\documentclass[prd,twocolumn,nofootinbib,superscriptaddress]{revtex4-1}
\usepackage{natbib}  \usepackage{bm}
\usepackage{epsfig} \usepackage[T1]{fontenc} \usepackage[latin9]{inputenc}
\usepackage[fleqn]{amsmath}  \usepackage{graphicx}
\usepackage{amssymb}

 \def\mnras{M.N.R.A.S}
\def\apj{Astrophys.J.} \def\apjl{Astrophys.J.Lett.}
\def\apjs{Astrophys.J.Supp.} \def\nat{Nature} \def\na{New Ast.}   \def\prl{Phys. Rev. Lett.}      \def\prd{Phys.Rev.D}
\def\araa{Annu.Rev.Astron.Astrophys.} \def\aap{Astron.Astrophys.}
\def\aaps{Astron.Astrophys.Supp.} 
 \def\Mesz{M\'esz\'aros~}

\begin{document}

\title{High Energy Neutrinos from Dissipative Photospheric Models of\\ Gamma
Ray Bursts } 
\author{Shan Gao} \email[Email: ]{sxg324@psu.edu}
\affiliation{Department of Physics, Department of Astronomy and Astrophysics,
Center for Particle Astrophysics, The Pennsylvania State University, University
Park, 16802, USA}
\author{Katsuaki Asano} \email[Email: ]{asano@phys.titech.ac.jp}
\affiliation{Interactive Research
Center of Science, Tokyo Institute of Technology, 2-12-1 Ookayama, Meguro-ku,
Tokyo 152-8550, Japan} 
\author{Peter \Mesz} \email[Email: ]{pmeszaros@astro.psu.edu}
\affiliation{Department of Physics, Department of Astronomy and Astrophysics,
Center for Particle Astrophysics, The Pennsylvania State University, University
Park, 16802, USA} \date{\today}

\begin{abstract} We calculate the high energy neutrino spectrum from gamma-ray
bursts where the emission arises in a dissipative jet photosphere determined by
either baryonically or magnetically dominated dynamics, and compare these
neutrino spectra to those obtained  in conventional internal shock models. We
also calculate the diffuse neutrino spectra based on these models, which appear
compatible with the current IceCube 40+59 constraints. While a re-analysis
based on the models discussed here and the data from the full array would be
needed, it appears that only those models with the most extreme parameters are
close to being constrained at present.  A multi-year operation of the full
IceCube and perhaps a next generation of large volume neutrino detectors may be
required in order to distinguish between the various models discussed.

\end{abstract} \maketitle


\section{Introduction} \label{sec:intro}

Gamma-ray bursts (GRBs) are a potential source of astrophysical high energy
neutrinos which are currently being investigated with IceCube. The standard GRB
internal shock scenario of neutrino production
\cite{Waxman+97grbnu,Guetta+04grbnu} used so far to compare against the IceCube
40 string and 40+59 string observations \cite{Abbasi+11-ic40diff,
Abbasi+12grbnu-nat} assumed some simplifications in the neutrino physics. Also,
the astrophysical model itself of the GRB prompt gamma-ray emission based on
the same internal shocks has been the subject of discussions in the gamma-ray
community \cite{Ghisellini+99grbspec,Preece+00batse,
Medvedev+04magshock,Meszaros06grbrev}, due to issues with the radiation
efficiency and the spectral properties in the standard version of this internal
shock scenario. 

For this reason, modified internal shock models that address these issues (e.g.
\citep{Asano+09slowheat,Inoue+11shockturb,Murase+12reac,ZhangICMART}) as well as
alternative models where the prompt gamma-ray emission arises in the jet
photosphere have been considered (e.g.
\citep{Meszaros+00phot,Ryde05,Rees+05phot,Peer+06phot,Beloborodov10pn,
Peer11-fermigrb}. In such photospheric models the high radiative efficiency is
due to dissipation processes in it. 

A separate question that has also been the subject of debate in the
astrophysics community is whether the jets in such relativistic sources are
dominated by the baryons or by magnetic fields, which imply different
macroscopic acceleration rates, different proper densities in the jet
rest-frame, and implying a major role for magnetic dissipation in the process
of particle acceleration. Such magnetically dominated jets in GRBs have been
investigated by \cite{Drenkhahn02,Tchekhovskoy+10grb,Metzger+11grbmag,
McKinney+12magphot} and others, and the gamma-ray emission is ascribed in such
models, again, to dissipative processes mainly in the photosphere, e.g.
\cite{Giannios+07photspec,Veres+12mag}. 

It is unclear at present whether the above mentioned modified internal shocks
or the photospheric models are best for interpreting the prompt gamma-ray
emission, nor whether the jet dynamics is dominated by baryonic or magnetic
stresses (e.g.  \cite{Meszaros+12raa}).  However, the expected neutrino
emission  is strongly dependent on the specific overarching dissipation and
dynamic model of GRBs.  For this reason, here we explore the neutrino features
of the three main types of models which are currently under consideration. We
illustrate the variety of astrophysical uncertainties involved in these models,
and how these can affect the expectations for detection with IceCube or future
instruments.


\section{The Dissipative Photospheric Scenarios} \label{sec:photosphere}

In the typical GRB model a high energy-to-mass ratio, jet-like relativistic
outflow is launched which initially accelerates with a bulk Lorentz factor
$\Gamma$, averaged over the jet cross section, whose dependence on distance
from the center of the explosion can be parametrized as 

\begin{equation} \Gamma=(r/r_{0})^{\mu}.  \label{eq:gamma} \end{equation} 

This behavior is
assumed valid up to a saturation radius $r_{\rm sat}= r_0\eta^{1/\mu}$, where the
Lorentz factor has reached the asymptotic value $\Gamma_{\rm sat} =\eta$, where
$\eta\simeq L/{\dot M}c^2$ is the dimensionless entropy of the outflow, $L$ and
${\dot M}$ being the average energy and mass flux. The index $1/3 \lesssim \mu
\lesssim 1$ ranges between the extreme $\mu=1/3$ magnetically dominated radial
outflow and the usual $\mu=1$ baryonically dominated outflow regimes, e.g.
\citep{Meszaros+11gevmag}. In the extreme magnetic case and baryonic
cases, the saturation radius is given by

\begin{equation} r_{\rm sat}= \begin{cases}
\eta^{3}r_{0}\sim2.7\times10^{14}\eta_{300}r_{0,7} ~{\rm cm},&~~{\rm
for}~~\mu=1/3\\ \eta r_0\sim 3\times 10^9 \eta_{300}r_{0,7} ~{\rm cm},&~~{\rm
for}~~\mu=1.  \end{cases} \label{eq:Rsat} \end{equation}

In the dissipative photospheric scenario, a fraction of the outflow bulk
kinetic energy is converted into radiation energy via some dissipation
mechanism in the neighborhood of the photosphere\footnote{We do not specify a
particular dissipation mechanism; e.g. the sudden drop of photon density might
trigger magnetic reconnection as proposed by \citep{McKinney+12magphot}, or it
could be due to MHD turbulence \citep{Thompson94} or shocks
\citep{Rees+05phot}, etc. We simply assume that as far as the detectable
radiation the dissipation around the photosphere plays the largest role, and we
concentrate on the photospheric dissipation region}., giving rise to the
``prompt" photon luminosity $L_{\gamma}=\epsilon_{e}L_{\rm tot}$, where $L_{\rm
tot}=10^{53}L_{53}$ erg/s is the isotropic equivalent total luminosity of the
jet\footnote{We also assume that the dissipation gives rise to a prompt photon
spectrum of the observed Band function type, without specifying the underlying
mechanism, e.g. seed photons scattered by electrons associated with turbulent
Alfven waves, synchrotron radiation from Fermi-I accelerated electron or
collisional mechanism by decoupled proton and neutron \citep{Thompson94,
Giannios+07photspec,Beloborodov10pn,McKinney+12magphot}}. The photospheric
radius is estimated by setting the Thomson optical depth
$\tau_{{\gamma}e}\sim{n^{\prime}_{e}}\sigma_{T}R_{\rm ph}/\Gamma=1$, where
$n^{\prime}_{e}=n^{\prime}_{p}\approx L_{\rm tot}/4\pi R_{\rm ph}^{2} m_{p}c^3
\eta \Gamma$ is the comoving density of electrons if the $e^{+}e^{-}$ pairs are
absent\footnote{The presence of pairs will increase the radius of the
photosphere by a factor of a few in a magnetized photosphere
\citep{Veres+12mag, Bosnjak+12delay} or in a baryonic photosphere where
dissipation is via MHD turbulence, or by a factor $\sim 20-30$ if baryonic
dissipation is via $pn$ colissions \citep{Beloborodov10pn}. In \S
\ref{sec:diffuse} we exemplify the effects of the effective photospheric radius
being larger.}. By using Eqn.\ref{eq:gamma} and the condition above, we obtain

\begin{eqnarray} \frac{R_{\rm ph}}{r_0}&=&\left(\frac{L_{\rm
tot}\sigma_T}{{4\pi} m_p c^3r_0}\right)\frac{1}{\eta \Gamma_{ph}^2} = \nonumber
\\ &=&\left\{ \begin{array}{ll} \eta_T^{1/\mu}(\eta_T/\eta)^3 &       {\rm if~
} \eta<\eta_T \\ \eta_T^{1/\mu}(\eta_T/\eta)^{1/(1+2\mu)}       &       {\rm
if~ } \eta>\eta_T \label{eq:Rph} \end{array} \right.  \end{eqnarray}

\citep{Veres+12mag}, where

\begin{equation} \eta_{T}=\left(\frac{ L_{\rm tot}\sigma_T}{4\pi m_p c^3
r_0}\right)^{\frac{\mu}{1+3\mu}}. \label{eq:etat} \end{equation} 

Typically, for a magnetically dominated $\mu=1/3$ case the photosphere occurs
in the acceleration phase $r\leq r_{\rm sat}$, if $\eta>\eta_T$, where
$\eta_T\simeq 150L_{53}^{1/6} r_{0,7}^{-1/6}$. On the other hand the
photosphere occurs in the coasting phase $r>r_{\rm sat}$ for $\eta<\eta_T$,
which is typical for baryonic cases, where $\mu=1$ and $\eta_T\simeq
1900~L_{53}^{1/4} r_{0,7}^{-1/4}$.  The Lorentz factor of the photosphere
$\Gamma_{\rm ph}$ has an $r_{\rm ph}$ dependence for $\eta>\eta_T$, being
$\Gamma_{\rm ph}=(r_{\rm ph}/r_0)\propto L^{\mu/(2\mu+1)} \eta^{-\mu/(2\mu+1)}
r_0^{-\mu/(2\mu+1)}$, while $\Gamma_{\rm ph}\approx\eta$ in the case
$\eta<\eta_T$.

In the baryonic photospheres the dissipation may be due to dissipation of MHD
turbulence \citep{Thompson94} or it may occur in the form of semi-relativistic
shocks \cite{Rees+05phot} with Lorentz factor $\Gamma_r \sim 1$, of different
kinematic origin but similar physical properties as internal shocks, with a
mechanical dissipation efficiency $\epsilon_d$. These result in a proton
internal energy, and result also in random magnetic fields with an efficiency
$\epsilon_B$, relativistic protons with $\epsilon_p$, and relativistic
electrons with $\epsilon_e$\footnote{An alternative baryonic dissipation
involves $pn$ collisions \citep{Beloborodov10pn} (see also
\citep{Bahcall+00pn}); here for simplicity and for intercomparison with other
models we just assume shock dissipation in the photosphere, whose effects are
comparable to those of magnetic dissipation.}.  In the magnetically dominated
jets the total jet luminosity $L_{\rm tot}$ in the acceleration phase before
dissipation occurs consists of a toroidal magnetic field component and a proton
bulk kinetic energy component. In the dissipation region a fraction
$\epsilon_d$ of $L_{\rm tot}$ is assumed to be dissipated, consuming a fraction
from each of the toroidal field and bulk proton energy, and resulting in proton
internal energy and in a fraction $\epsilon_B$ which appears as random magnetic
fields, and $\epsilon_p$ and $\epsilon_e$ which appear as relativistic protons
and relativistic electrons.  In both baryonic and magnetically dominated cases
we assume $\epsilon_B+\epsilon_p+\epsilon_e=1$, and we take $\epsilon_{d}\sim
0.3$ and $\epsilon_B\sim 1/3$ as examples in this paper. In the jet comoving
frame the random magnetic field after the dissipation is parametrized by an
energy density

\begin{equation} U_{\rm B,random}^{\prime}=B^{\prime2}/{8\pi}=
\kappa\epsilon_{B}\epsilon_{d}L_{\rm tot}/(4\pi{R_{\rm ph}^{2}}\Gamma_{\rm
ph}^{2}c) \label{eq:Bfield} \end{equation}

where for semi-relativistic shocks ($\Gamma_{\rm rel}\sim 1$),  and a
compression ratio of $\kappa\sim4$ is assumed. For the magnetically dominated outflow,
during the acceleration phase, the energy remaining in toroidal fields after
dissipation is 

\begin{equation} U_{\rm B,toroid}^{\prime}=(1-\epsilon_{d})(1-\Gamma_{\rm
ph}/\eta)L_{\rm tot} /(4\pi{R_{\rm ph}^{2}}\Gamma_{\rm ph}^{2}c) \end{equation}

where $\Gamma_{\rm ph}$ is the bulk Lorentz factor of the protons at the
photosphere.

Calculations and simulations of of such baryonic and magnetic dissipative 
photospheres as well as internal shocks generally result in an escaping photon
spectrum similar to the observed characteristic ``Band" spectrum
\citep{Band+93}, parametrized as

\begin{equation} dN_{\gamma}/dE\propto(E/E_{\rm br})^{x_{\rm ph}}
\label{eq:photon} \end{equation}

in the observer frame. Observationally, for average bursts at redshifts $z\sim
2$ the mean values are $E_{\rm br}\sim 300$ keV, $x_{\rm ph}=-1$ below $E_{\rm
br}$ and $x_{\rm ph}=-2$ above $E_{\rm br}$. In a photosphere this spectral
shape is the product of the modification of a thermal spectrum by the
dissipation.  For the purposes of this article, we treat this photon spectrum
as the input for our calculations, transformed to the rest frame of the
outflow. While the bulk Lorentz factors in the photosphere and internal shock
models may differ, for the purposes of comparison we adopt here as a test case
the same comoving frame photon spectral break energy for the dissipation zones
of the various models considered, $E_{\rm br}^{\prime}=0.01$ MeV, and $x_{\rm
ph}=-1$ below $E_{\rm br}^{\prime}$ and $x_{\rm ph}=-2$ above 
$E_{\rm br}^{\prime}$.   The lower
and upper branches can have cut-off energies, e.g.  determined by synchrotron
self-absorption below and acceleration restrictions or $\gamma\gamma\to
{e^{+}e^{-}}$ pair production above, the cut-off values depending on the
specific model and its parameters.  For simplicity, here we adopt the same
constant values of a lower limit $E_{\rm min}^{\prime}=1$ eV and an upper limit
$E_{\rm max}^{\prime}=0.5$ MeV, which are adequate for our purposes since the
neutrino results are insensitive to these values. The total luminosity of this
Band-function spectrum is normalized to $\epsilon_{d}\epsilon_{e}L_{\rm tot}$,
where $L_{\rm tot}$ represents the total luminosity. (An additional softer
thermal spectral component can also be present at the photosphere.  However,
the temperature of this component is estimated as $T\sim{o(1)\times{\rm keV}}$
at the photosphere \citep{Veres+12mag}, corresponding to a thermal luminosity
$L_{\rm thermal}\sim{o(1)}\times10^{49}{\rm erg/s}$ which is low compared to $L_{\rm
tot}$ and $L_{\gamma}$. Hence we have neglected this component for the purposes
of the present neutrino calculation.)

When the outflow encounters the external medium, it starts to decelerate at a
radius \begin{align} R_{d}& \approx(\frac{3L_{\rm tot}t_{\rm dur}}{4{\pi}n_{\rm
ISM}m_{p}c^2\eta^{2}})^{1/3} \nonumber \\ & =2.54\times10^{16}L_{t,53}^{1/3}
(t_{\rm dur}/10s)^{1/3}n_{\rm ISM,2}^{-1/3}\eta_{300}^{-2/3} \label{eq:rd}
\end{align}

where an external shock forms which is also able to produce neutrinos. Here we
have assumed a uniform interstellar medium of particle density $n_{\rm
ISM}=10^{2}n_{\rm ISM,2} cm^{-3}$ and a jet outflow duration time $t_{\rm dur}$
in the central engine frame. The interstellar density value does not affect the
photospheric or internal shock neutrinos, but it does affect the external shock
neutrinos. Here we have adopted a density which is optimistic for the external
shock neutrinos, since even so the external neutrino fluxes predicted are low
and more moderate densities such as the typically used $n_{\rm ISM}=1 \rm
cm^{-3}$ would lead to even smaller external shock neutrino fluxes. The
corresponding deceleration timescale is estimated as
$t_{d}^{\prime}\sim{R_{d}}/c\eta$. At this radius deceleration $R_d$ the
external shock has fully developed, consisting of a forward shock, and possibly
also a reverse shock (if the magnetization parameter $\sigma$ is or has become
low enough at this radius). If present, for our parameters the reverse shock is
marginally in the so-called thin-shell regime, the reverse shock having become
semi-relativistic as it crosses the ejecta at about the deceleration time
$t_{d}$.

\begin{figure} \includegraphics[width=0.99\columnwidth]{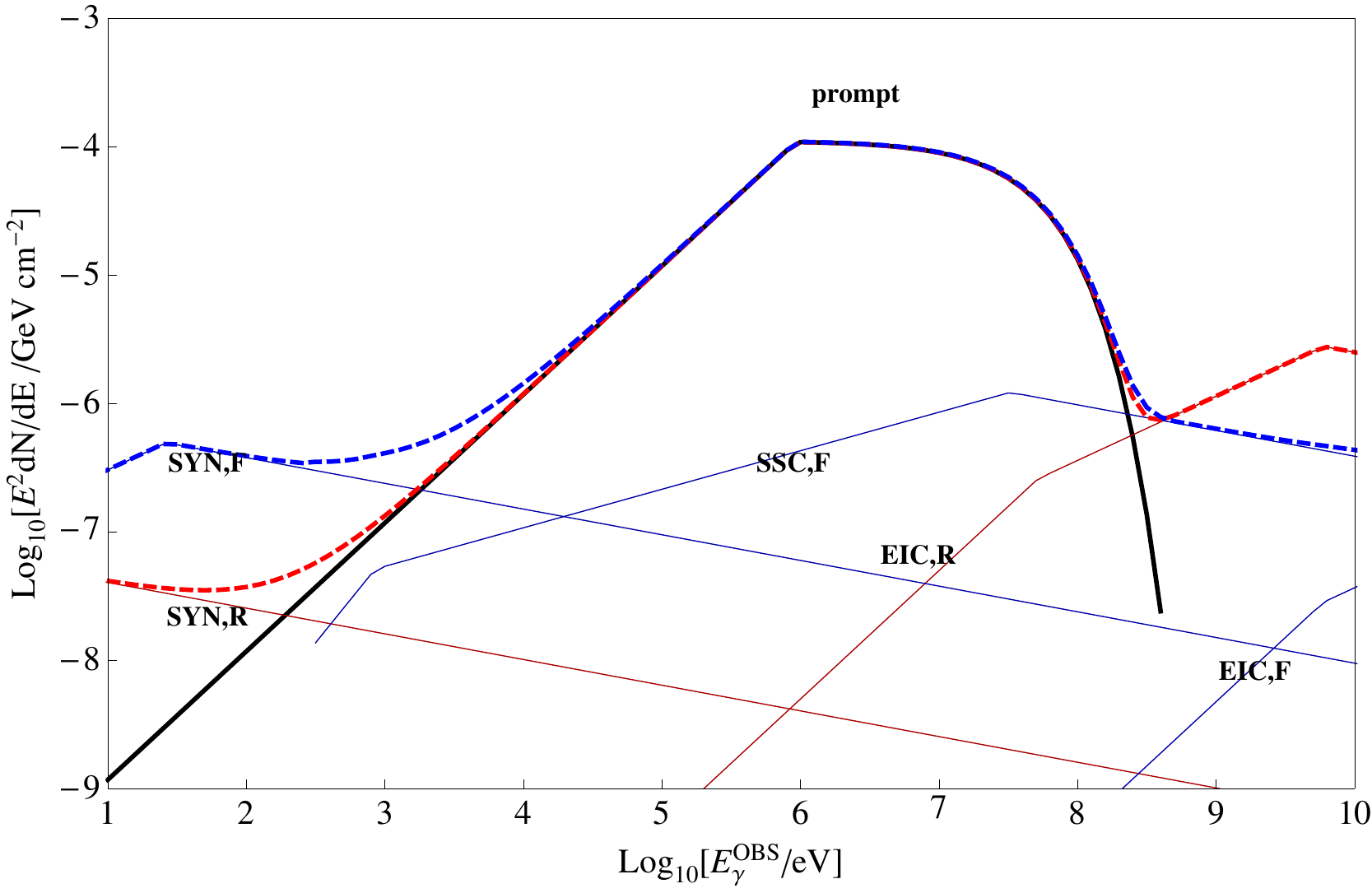} \caption{
Photon spectrum in the observer frame for a typical GRB with parameters $L_{\rm
tot}=5\times10^{52}$ erg/s, $t_{\rm dur}=10$ s, $\eta=300$ , $z=1.0$, $n_{\rm
ISM}=100~\rm cm^{-3}$ , $\epsilon_{\rm d}=0.3$, $\epsilon_{\rm e}=\epsilon_{\rm
p}=\epsilon_{\rm B}=1/3$, where $\epsilon_{\rm d}$ is defined as the total
dissipated energy from jet total energy, in the forms of $\epsilon_{\rm
e}$,$\epsilon_{\rm p}$ and $\epsilon_{\rm B}$. The subindex "ph" refers to
photosphere, ``prompt" refers to the prompt emission from the photospheric
region, for which a Band-like spectrum is assumed.  The SYN,R and SYN,F are the
synchrotron from the reverse and forward shock; SSC is the synchrotron-self
Compton spectrum, EIC is the inverse Compton scattering of the prompt photons
in the external shock region.  (At very high energies, a Klein-Nishina break
may be expected; however, for the neutrino calculation this contribution can be
neglected since it contains very few photons, hence these KN breaks are not
shown here). A smaller dissipation value for external shock emission used ,
$\epsilon_{\rm B,FS}=\epsilon_{\rm B,RS}=0.02\epsilon_{\rm B}$, 
suitable to explain the external shock photon emission \citep{Veres+12mag}.  } 
\label{fig:1} \end{figure}

The turbulent magnetic fields generated in these external shocks lead to
synchrotron radiation, as well as synchrotron self-Compton (SSC) and external
inverse Compton (EIC) scattering of non-thermal photons from the dissipation
region near the photosphere or the internal shocks. The detailed method of
calculation of these photon spectra are discussed in
\citep{Veres+12mag,Toma+11pop3,Asano+11fermiextra}.  As shown below, however,
the neutrino fluence from the external shock region is a few orders of
magnitude lower than that from the photospheric or baryonic internal shock
regions, due to a much lower photon density leading to a lower interaction rate
and lower pion production efficiency. Under the assumptions made here, the
input photon spectrum of a GRB with typical parameters is shown in
Fig.\ref{fig:1} as an example\footnote{The photospheric spectrum here does not
include the effect of the relativistic leptons injected if we had included
nuclear collisions \cite{Beloborodov10pn}; the effect would be to extend the
upper branch of the Band photon spectrum into the GeV range; however, it is the
photons around the Band peak that affect significantly the photo-pion neutrino
production discussed here.}.  For the reverse shock, we include both the
self-generated photons from the reverse shock and the prompt emission as target
photons for inverse Compton scattering as well as for $p\gamma$ interaction.
For the forward shock, we include the forward shock (FS), reverse shock (RS)
and prompt photons.


\section{method of calculation} \label{method}

In the baryonic dissipation regions, whether these are in the photosphere or in
internal shocks beyond the photosphere, it is usually assumed that protons, as
well as electrons, are accelerated through a Fermi-first order (Fermi-I) acceleration
mechanism in the disordered magnetic fields created in the region. A process
similar to Fermi acceleration is also expected in magnetic reconnection regions
where layers of magnetic field of opposite polarity meet and drive converging
flows.  \citep{Kowal+09FastRC,Sironi+11,Hoshino+12}.  The particles bounce back
and forth in the converging flow between the layers and can reach similar
maximum energies as in the usual Fermi mechanism. We assume that the injection
of accelerated protons has a spectrum

\begin{equation} dN_{p}/dE\propto{E}^{-x_{p}} \hspace{5 mm} (E_{\rm
p,min}<E_{\rm p}<E_{\rm p,max}) \label{eq:pspectrum} \end{equation}

with $x_{p}=2$ as a nominal value.  The acceleration timescale is $t_{\rm
acc}^{\prime}\sim{t_{\rm cycle}^{\prime}}= \xi_{p}r_{g}^{\prime}/c=
\xi_{p}E_{p}^{\prime}/eB^{\prime}c$ , where $r_{g}$ is the average gyroradius.
Here we have adopted a minimum injection energy of protons $E_{\rm
p,min}^{\prime}=10$ GeV since the neutrino spectrum is insensitive to this
value. We have also assumed a high compression ratio and weakly disordered
magnetic fields, corresponding to $\xi_{p}\sim10$. The accelerated proton
spectral energy (eqn.\ref{eq:pspectrum}) here is normalized to a fraction of
the jet total luminosity $\epsilon_{p}L_{\rm tot}$, the value of $\epsilon_{p}$
being discussed in section \ref{sec:single}.

The maximum proton energy is constrained by the gyroradius being smaller than
the size of the acceleration region $r_{g}^{\prime}<R_{\rm ph}/\Gamma_{\rm
ph}$, or by radiative cooling $t_{\rm p,acc}<t_{\rm p,cool}$ where $t_{\rm
p,cool}$ is the total cooling timescale for the proton $t_{\rm
p,cool}^{-1}=t_{\rm p\gamma}^{-1}+t_{pp}^{-1}+t_{BH}^{-1}+t_{sy}^{-1}+
t_{IC}^{-1}+t_{ad}^{-1}$. The terms on the right hand side are the
photohadronic, $pp$ collisional, Bethe-Heitler (photopair), proton synchrotron,
inverse Compton and adiabatic inverse cooling timescales in the fluid comoving
frame (we have dropped the "prime" superscript here), given respectively by

\begin{align} t_{p\gamma}^{-1}&
{\approx}\frac{c}{2\gamma_{p}^{2}}\int_{0}^{\infty}\frac{dE}{E^{2}}n_{ph}(E)
\int_{\epsilon_{\rm TH}}^{2\gamma_{p}E}d{\epsilon}
\epsilon{\sigma_{p\gamma}}(\epsilon)K_{p\gamma}(\epsilon) \label{eq:pg}\\
t_{pp}^{-1}&=cn_{p}\sigma_{pp}(\gamma_{p})K_{pp}(\gamma_{p}) \label{eq:pp}\\
t_{BH}^{-1}&\approx\frac{7(m_{e}c^{2})^{2}\alpha_{f}\sigma_{T}c}
{9\sqrt{2}{\pi}m_{p}c^2\gamma_{p}^{2}}
\int_{{\gamma_p}^{-1}}^{\infty}d\gamma_{e}\gamma_{e}^{-2}n_{ph}(\gamma_{e}m_{e}c^2)
\nonumber \\ &\times
\{(2\gamma_{p}\gamma_{e})^{3/2}[\log(2\gamma_{p}\gamma_{e}) -2/3]+2/3\}
\label{eq:BH}\\
t_{sy}^{-1}&=4\sigma_{T}m_{e}^{2}\gamma_{p}(B^{2}/8\pi)/3m_{p}^{3}c\label{eq:psy}\\
t_{IC}^{-1}&=\frac{3(m_{e}c^2)^{2}\sigma_{T}c}
{16\gamma_{p}^{2}(\gamma_{p}-1)\beta_{p}}
\int_{0}^{\infty}\frac{dE}{E^{2}}F(E,\gamma_{p})n_{ph}(E) \label{eq:pIC}\\
t_{ad}^{-1}&\approx{\Gamma{c}/R}\label{eq:adb} \end{align}

where each individual cooling inverse timescale is defined as
$t^{-1}\equiv-(d\gamma_{p}/dt)\gamma_{p}$ and $n_{ph}(E)\equiv{dN/dEdV}$ is the
photon differential spectral density.  Neutrinos result mainly from charged
pion and kaon decays, to the first and second leading order of approximation
respectively here. These charged mesons come from $p\gamma$ and $pp$
interactions (eqn.\ref{eq:pg},\ref{eq:pp}). The cross section and inelasticity
in the former channel, considering the lower threshold $\epsilon_{\rm TH}=0.2$
GeV, are approximated by two step-functions:

\begin{align} \sigma_{p\gamma}(\epsilon) & =\begin{cases} 340{~\mu}b &
0.2<\epsilon/GeV<1.0\\ 120{~\mu}b & \epsilon>1.0 \end{cases}\\
K_{p\gamma}(\epsilon) & =\begin{cases} 0.2 & 0.2<\epsilon<1.0\\ 0.6 &
\epsilon>1.0 \end{cases} \end{align}

where $\epsilon$ is the photon energy in the proton comoving frame. For
$0.2<\epsilon<1.0$ GeV the cross section is dominated by resonances while for
$\epsilon>1.0$ GeV multi-pion production takes over. In the single-pion
resonance channel, $\pi^{+}$ and $\pi^{0}$ are created at approximately the
same rate. In the multi-pion channel, we assume that pions are created with an
average multiplicity of 3, and in a first-order approximation the $\pi^{+}$ ,
$\pi^{-}$ and $\pi^{0}$ come in equal numbers.  (For more details, see e.g.
\citep{Dermer+09book}).  For $pp$ interactions, the (thermal) protons which are
not accelerated in the fluid are the targets. In the fluid comoving frame,
where the target protons are basically at rest, the incident proton with energy
$E_{p}(E_{p}\gg1$ GeV) has a cross section approximated by

\begin{equation} \sigma_{\rm pp}{\approx}\sigma_{\rm
pp,inel}\approx30[0.95+0.06\log(E_{p})] {\rm ~mb} \end{equation}

with charged pion multiplicity approximated by

\begin{equation} \chi(s){\approx}1.17+0.3\log{s}+0.13\log^{2}s \end{equation}

in which $s$ is the invariant energy $\sqrt{s}$ squared of the binary particle
system. The detailed pion spectra are discussed in e.g.
\citep{Kamae+06,Gao+12magnu}. However, in this paper we take the approximation
that the pions are created at rest in the CM frame of the binary particle
system. The error caused by this approximation is much reduced in the
calculation of a broad spectrum in the high energy regimes (compared with the
case in \citep{Kamae+06} or \citep{Gao+12magnu}),and is smaller than the
astrophysical uncertainties in this paper.

\begin{figure} \includegraphics[width=0.99\columnwidth]{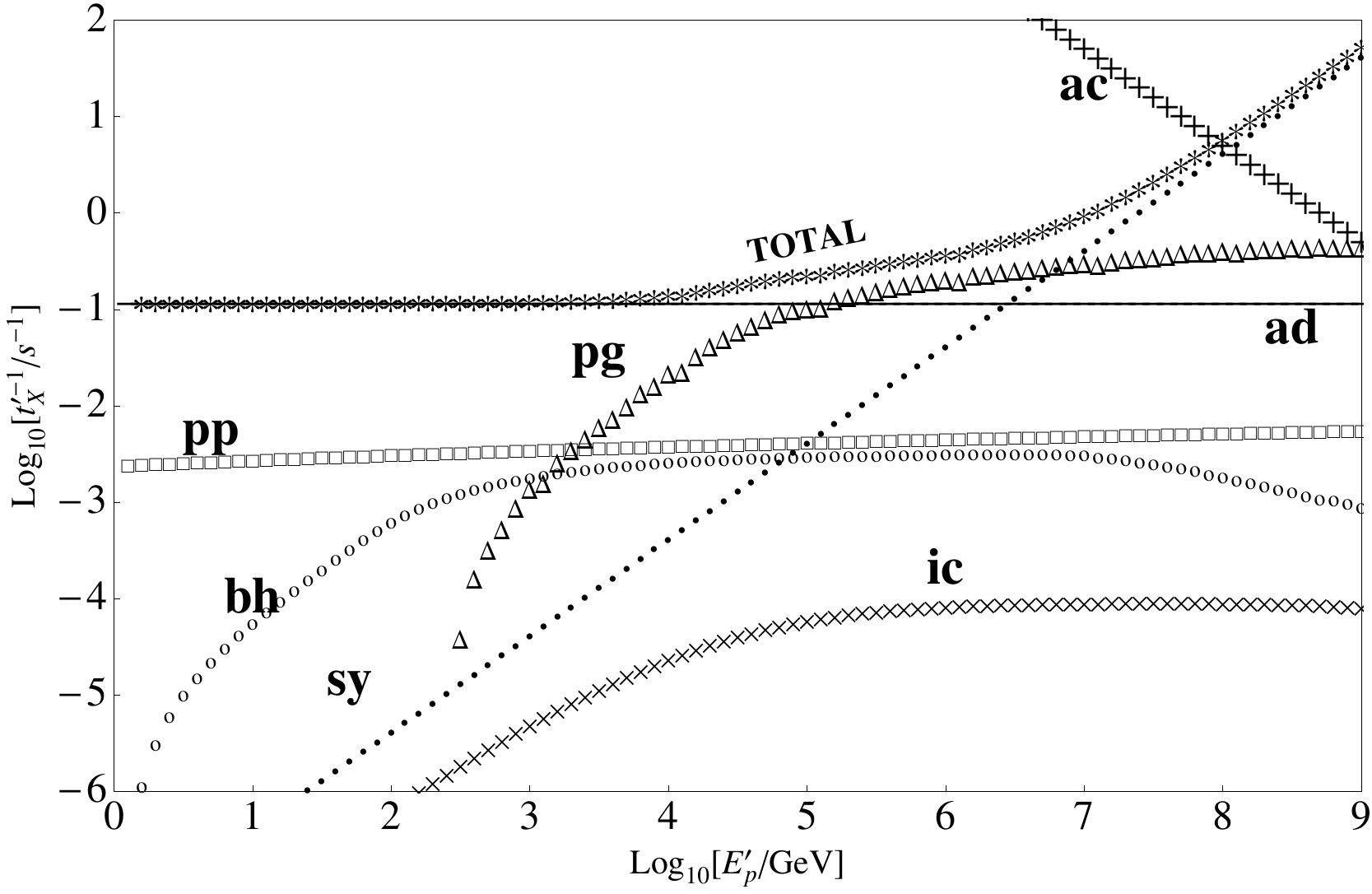} \caption{Proton
inverse cooling timescales at the photosphere as a function of proton energy in
the magnetic reconnection region comoving frame, defined as $t^{-1}=|E dt/dE|$
in s$^{-1}$.  The symbols used are Triangle(pg): $p$-gamma (photopion)
interaction;  Circle(bh): Bethe-Heitler (photopair); Square(pp): $pp$
interaction;  Dot(sy): proton synchrotron;  Cross (ic) : proton inverse
Compton;  Solid line(ad): adiabatic cooling;  Plus(ac) : Fermi acceleration
timescale; Minus(TOTAL) : total cooling timescale.  The astrophysical
parameters for this GRB are $L_{\rm tot}=10^{53.5}$ erg/s, $\eta=300$. At lower
energies, protons are mainly cooled by adiabatic expansion and $pp$ collisions.
At intermediate energies in this figure, protons are mainly cooled by photopion
production and at higher energies by proton synchrotron radiation.}
\label{fig:2} \end{figure}

To derive eqn.\ref{eq:BH}, we have used a cross section
$\sigma_{\phi{e}}\approx(7/6\pi)
\alpha_{f}\sigma_{T}\log(\epsilon_{\gamma}/2m_{e}c^2)$ where $\alpha_{f}=1/137$
is the fine structure constant, $\sigma_{T}=665\rm mb$ is the Thomson cross
section and $\epsilon$ is the photon energy in the proton rest frame. We note
that although $p\gamma{\to}p{e^{\pm}}$ has a larger cross section than the
photopion process, the effective inelasticity of the proton is smaller and the
relative photopion and photopair energy loss rate for protons interacting with
the peak of the $\nu{F}_{\nu}$ target photon spectrum is
$K_{\phi\pi}\sigma_{\phi\pi}/K_{\phi{e}}\sigma_{\phi{e}}\approx100$.  An
expression for the function $F(E,\gamma_{p})$ in eqn.\ref{eq:pIC} is given by
\citep{Jones+65}. With the above analytical approximate expressions we can
calculate the energy fraction from the parent proton spectrum converted into
pions, $f_{p\gamma}{\equiv}t_{\rm p\gamma}^{-1}/t_{\rm p,cool}^{-1}$ and
$f_{pp}{\equiv}t_{pp}^{-1}/t_{\rm p,cool}^{-1}$. With the leading order
approximation that the pions are created at the rest frame of the protons, we
can obtain the pion spectrum. We also rouphly approximate the produced Kaon
number density as $1\sim10$\% of the pions from $p\gamma$ or $pp$ interactions,
motivated by \citep{Asano+06,Ando+05} or simulations using PYTHIA-8. Neutrinos
from Kaon decays are generally subdominant but they become the main component
at the high end of the neutrino spectrum. At these energies charged Kaons
suffer less from radiative cooling than charged Pions due to their larger mass
and shorter lifetime.  The various cooling timescales for a GRB with typical
parameters are shown in Fig.\ref{fig:2}.


The pion decay kinematics are well established. Neutrinos result mainly from
the following channel:

\begin{equation}
\pi^{\pm}\rightarrow\mu^{\pm}+\nu_{\mu}(\bar{\nu}_{\mu})\rightarrow
e^{\pm}+\nu_{e}(\bar{\nu}_{e})+\nu_{\mu}+\bar{\nu}_{\mu}~.  \label{eq:piontonu}
\end{equation}

which we calculate in detail following the method in \citep{Marscher+80}. A
high energy pion may lose a significant fraction of its energy through
synchrotron radiation before it decays. Therefore we first calculate the pion
spectrum after the cooling process has set in (using a method similar to
eqns.[\ref{eq:psy},\ref{eq:pIC} and \ref{eq:adb}]). Then we calculate the
neutrino spectrum from the 'final' pion and muon spectra. The $\mu^{\pm}$ has a
longer mean life-time and smaller mass which makes its synchrotron cooling more
severe than that of charged pions. Finally we note that the leading decay
channel of the charged kaon is the same as that of the charged pion so in this
sense they can be viewed as "effective pions".

While the maximum proton energy at injection is determined by the condition
$t_{\rm acc} < t_{\rm cool}$, the maximum cosmic ray energy of the escaped
protons can be smaller, being given by the condition $t_{\rm cool} > t_{\rm
dyn}$. The resultant values are summarized in Table.\ref{tbl:1}.


\section{Parameters and Neutrinos from a Single Source} \label{sec:single}

We numerically compute the neutrino spectrum by using the method described in
the previous section. Several parameters are needed: the jet total luminosity
$L_{\rm tot}$ and its duration in the source frame $t_{\rm dur}$, the target
photon luminosity $L_{\gamma}=\epsilon_{d}\epsilon_{e}L_{\rm tot}$, the Fermi
accelerated proton luminosity and its power-law spectral index $x_{p}$, the
magnetic field calculated from the parameter $\epsilon_{B}$ and
eqn.\ref{eq:Bfield}, the dissipation radius $R$, the outflow energy to mass
ratio $\eta$ and finally the source redshift $z$.


\begin{table*} \includegraphics[width=1.9\columnwidth]{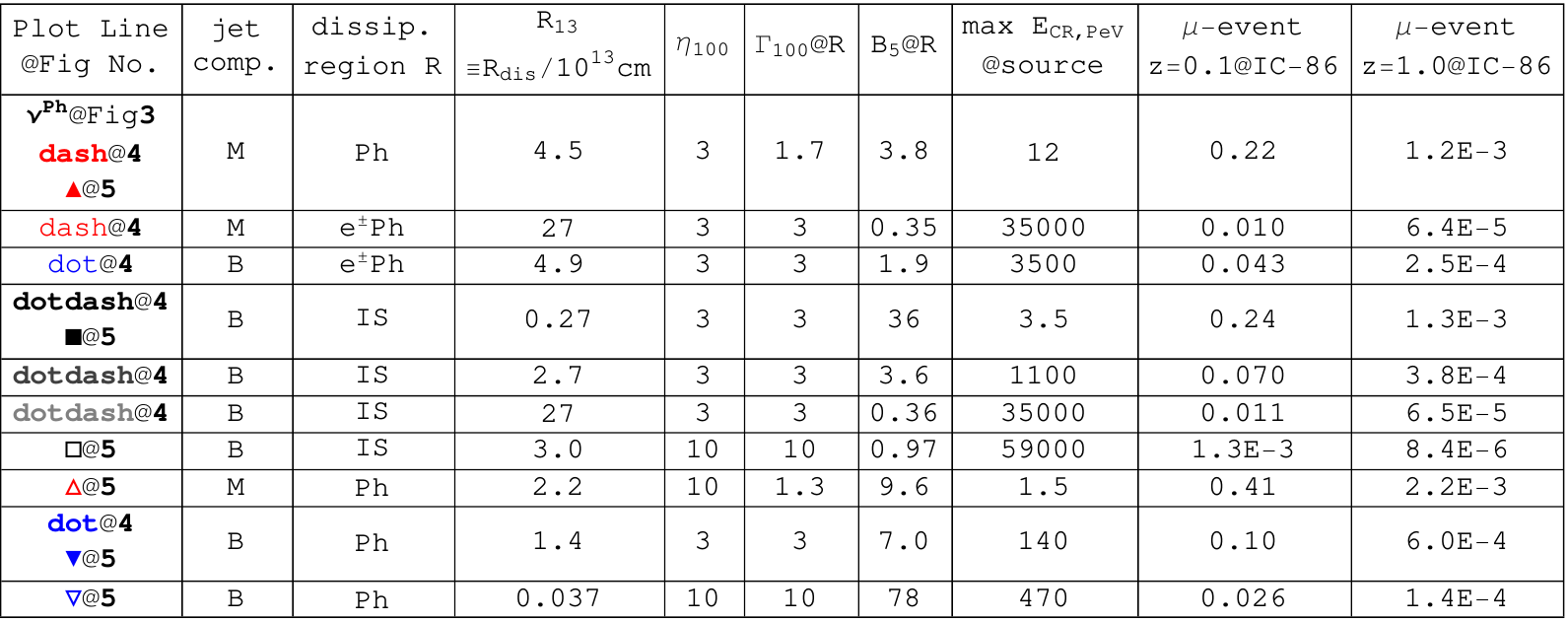} \caption{
Parameter list for different models calculated. The common parameters are: jet
total luminosity $L_{\rm tot}=10^{53.5}$ erg/s, source frame duration $t_{\rm
dur}=10$ s.  source redshift $z=0.1$, and the dissipation partition fractions
$\epsilon_{\rm e}=\epsilon_{\rm B}=\epsilon_{\rm p}=1/3$ with $\epsilon_{\rm
d}=0.3$ . The first column identifies the different curve symbols and the
figure used.  The M and B in the second column refer to "Magnetic dominated"
and "Baryonic dominated".  The third column identifies the type of dissipation
region: Ph for photosphere with $R_{\rm ph}$ from eqn.\ref{eq:Rph},
$e^{\pm}{\rm Ph}$ for pair-photosphere, and IS for internal shock of radius
$R_{\rm IS}$ from eqn.  \ref{eq:Ris}. In the next columns
$R\equiv10^{13}R_{13}$ cm gives the corresponding radii, $\eta_{100}$ is the
initial dimensionless entropy, $\Gamma_{100}@R$ and $B_{5}=B_5@R$ are the bulk
Lorentz factor and comoving magnetic field (in the unit of $10^{5}$ G) in the
dissipation region. The max $E_{\rm CR,PeV}$ is the maximum escaping cosmic ray
(proton) energy in the source frame, calculated by setting $t_{\rm
p\gamma}=t_{\rm dyn}$, or $t_{\rm p,syn}=t_{\rm dyn}$, or $t_{\rm
p,cool}=t_{\rm p,acc}$, whichever gives the smallest $E$.  The number of
$\mu$-events are in the last two columns; these are estimated from the neutrino
flux and the effective area of the IceCube 86-string configuration, for a
source at $z=0.1$ or $z=1.0$. } \label{tbl:1} \end{table*}

We discuss three main representative scenarios: an extreme magnetic photosphere
model where $\Gamma\propto{r}^{1/3}$, a baryonic photosphere model where
$\Gamma\propto{r}$, and a modified internal shock (IS) scenario where two
ejecta shells of different bulk Lorentz factors collide. For the modified
internal shocks we assume that a high mechanical dissipation efficiency
$\epsilon_d$ is achieved, e.g.
\citep{Asano+09slowheat,Inoue+11shockturb,Murase+12reac}, and that the photon
spectral issues raised about traditional internal shocks are avoided in such
mechanisms. As far as neutrino production, the location and seed photon
spectrum is similar to that in the standard internal shock, but without the
simplifications of \cite{Guetta+04grbnu, Abbasi+12grbnu-nat} in the neutrino
physics; that is, we treat the $p\gamma$ interaction with the whole photon
spectrum, not just the break region, and include besides the $\Delta$-resonance
also multi-pion effects, Kaons and detailed secondary particle distributions
for the charged meson and muon decay, as in e.g.
\cite{Hummer+11nu-ic3,Asano+12grbhad}. We assume for the internal shocks a
dissipation efficiency $\epsilon_d=$ 0.3, which for comparison is taken to be
similar to that of the photospheric models.  For the two photospheric models,
the extreme magnetic one satisfies $\eta >\eta_T$, while for the baryonic one
$\eta <\eta_T$, so the photospheric dissipation radii are

\begin{equation} \frac{R_{\rm ph}}{r_0} = \begin{cases}
\eta_T^{18/5}\eta^{-3/5} &~~{\rm for}~~ \mu=1/3 ;\cr \eta_T^4 \eta^{-3} &~~{\rm
for}~~ \mu=1~, \end{cases} \end{equation}

where $\eta_T$ is defined in eq. (\ref{eq:etat}).  For the internal shock
models, the dissipation radius  at which these shocks occur is

\begin{equation} R_{\rm IS}\approx{c}\eta^{2}t_{\rm var}.  \label{eq:Ris}
\end{equation}

\begin{figure} \includegraphics[width=0.99\columnwidth]{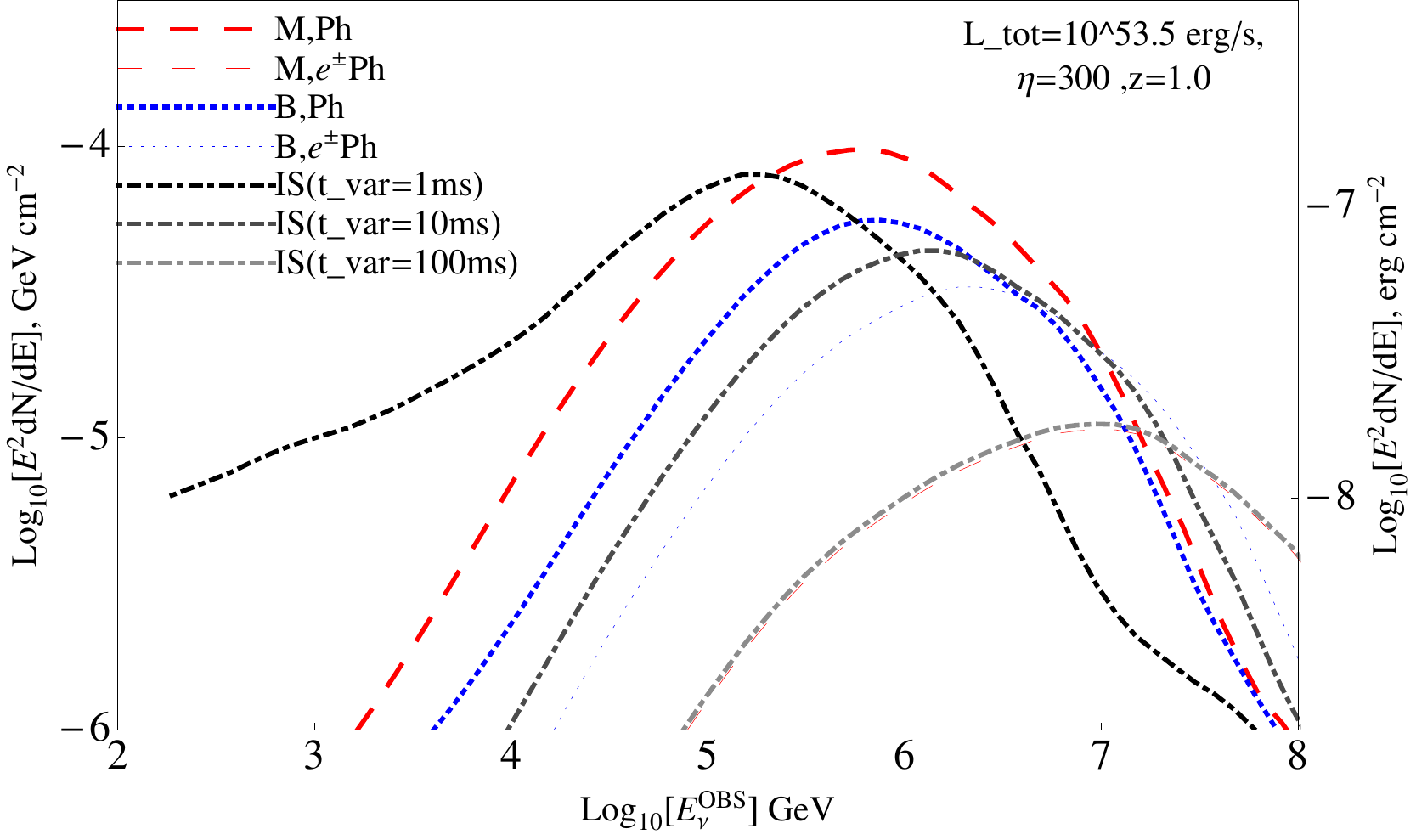} \caption{The
$\nu_{\mu}+\bar{\nu}_\mu$ fluence spectra for different models assuming a
source at redshift $z=1$. $L_{\rm tot}=10^{53.5}$ erg/s, $\eta=300$ are used.
(External shock neutrino spectrum is not calculated here.)   The models shown
here are the  $\mu=1/3$ magnetic photosphere (M,Ph), $\mu=1$ baryonic
photosphere (B,Ph) , pair-photosphere dissipation (M,$e^{\pm}{\rm Ph}$ or
M,$e^{\pm}{\rm Ph}$) and internal shock shock models (IS) with $t_{\rm
var}=1,10,100$ ms (black, dark gray, light gray) } \label{fig:4} \end{figure}

\begin{figure} \includegraphics[width=0.99\columnwidth]{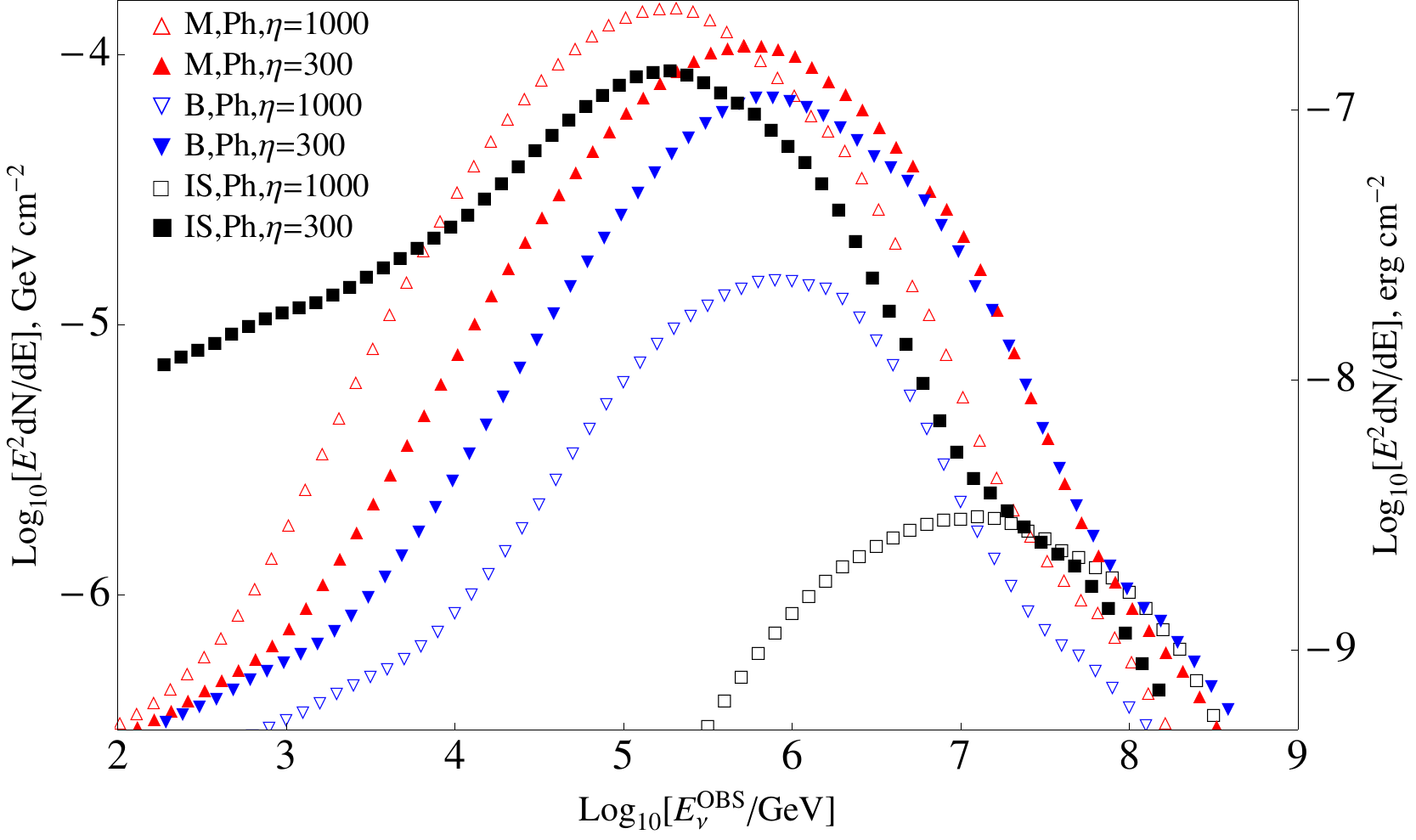} \caption{The
neutrino spectrum from a magnetic photosphere (red upstanding triangle), a
baryonic photosphere (blue inverted triangle) and an internal shock model
(black square), for $\eta=300$ (filled) $\eta=1000$ (empty). Both the two lines
for internal shock models have used $t_{\rm var}=1$ ms. The other parameters
are the same as in fig.\ref{fig:4}. Note that a slower jet produces a higher
neutrino flux in all these cases. The more detailed quantities for those models
(and in fig.\ref{fig:4}) are listed in Table.\ref{tbl:1}.  } \label{fig:5}
\end{figure}

Here $\eta$ is the average Lorentz factor of the two shells and $t_{\rm var}\gtrsim
1$ ms is the variability timescale which represent the time interval between
the ejection of the two shells in the source frame. Note that a range of
$t_{\rm var}$ is indicated by observations, extending down to $t_{\rm
var}=0.001\sim1$ s. This introduces a large uncertainty in the internal shock
radius and the corresponding final neutrino spectrum. Here we use optimistic
values for the internal shock neutrino production, for comparison purposes.
Unless specified otherwise, in the following we assume a nominal parameter set
of $L_{\rm tot}=10^{53.5}{\rm erg/s}$ , $t_{\rm dur}=10$ s (source frame) ,
$\eta=300$ , $\epsilon_{d}\epsilon_{e}=0.1$ corresponding to an isotropic
equivalent total photon luminosity of $L_{\gamma}\approx10^{52.5}$ erg/s (which
is roughly the average luminosity from GRB statistics).  The exact energy
partition fractions in the jet are not well known, here we have assumed
$\epsilon_{B}=\epsilon_{e}=\epsilon_{p}=0.33$, $\epsilon_{d}=0.3$.  As an
example of the fluence in the observer frame , we consider a GRB at redshift
$z=1$ corresponding to a luminosity distance of 6.6 Gpc, and at $z=0.1$
corresponding to a luminosity distance of 450 Mpc in a standard ${\rm
\Lambda{CDM}}$ cosmology with
 
\begin{equation} \Omega_{m}=0.28 , \Omega_{V}=0.72 , H_{0}=72 {\rm km/s/Mpc}.
\label{eq:LCDM} \end{equation} The detailed parameters of different models are
listed in Table.\ref{tbl:1} for which the neutrino spectra are plotted in figs.
\ref{fig:4} and \ref{fig:5}.

\begin{figure} \includegraphics[width=0.99\columnwidth]{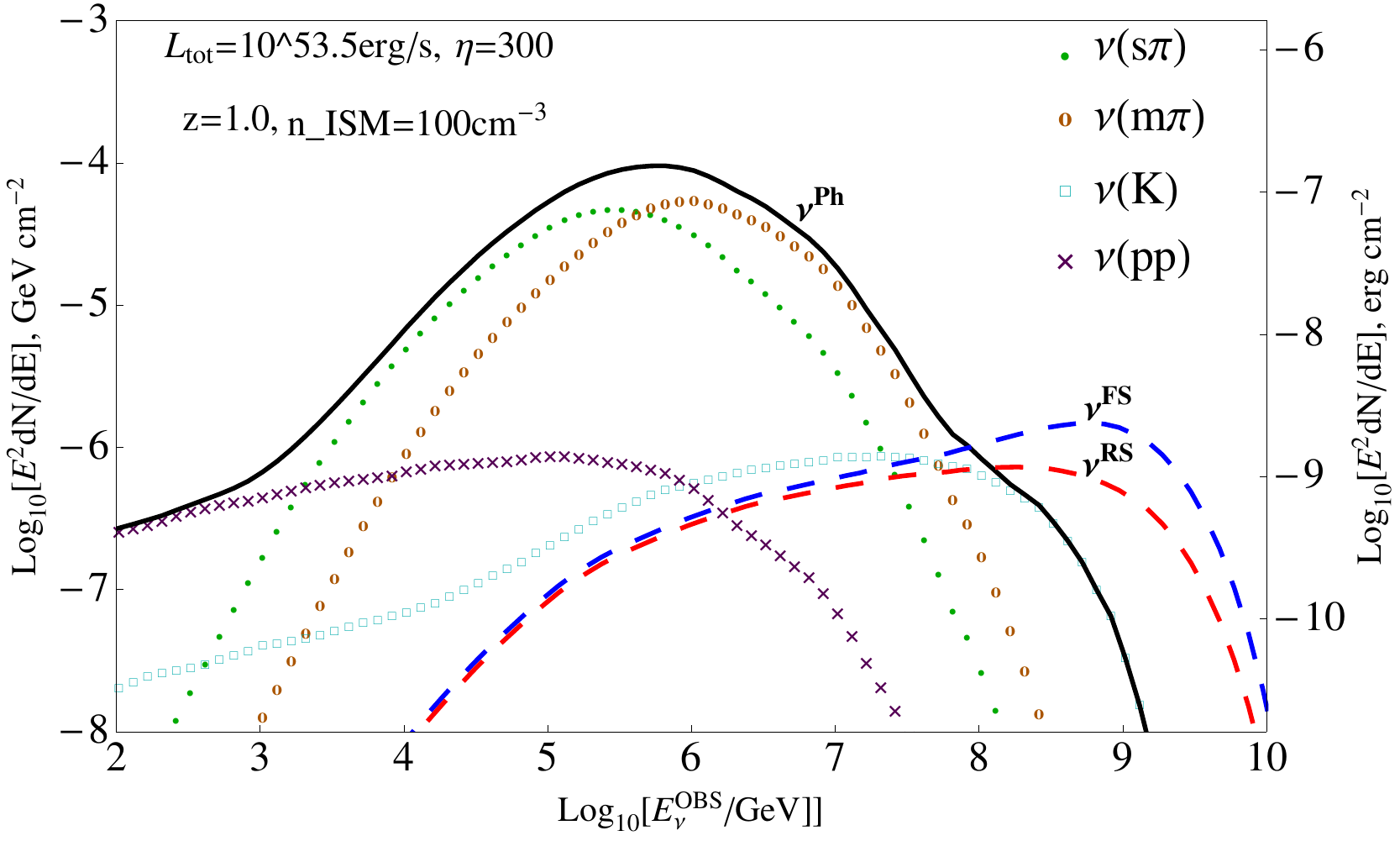} \caption{The
observer frame $\nu_{\mu}+\bar{\nu}_{\mu}$ fluence $E^{2}dN/dE$ for a $\mu=1/3$
magnetically dominated burst at $z=1$. Solid line($\nu^{\rm Ph}$): total muon
neutrino spectrum from $\Delta$-resonance (dotted), $p\gamma$ multi-pion
production (circle), $pp$ collision (cross) and Kaons(square).  Large dashed
line ($\nu^{RS}$ and $\nu^{FS}$): total neutrino spectrum from external reverse
and forward shock in the early afterglow. For the forward shock neutrino
calculation, we included the photons from the reverse shock as well.}
\label{fig:3} \end{figure}

The neutrino spectral fluence ($\nu_{\mu}+\bar{\nu}_{\mu}$) from a magnetically
dominated $\mu=1/3$ GRB is shown in Fig.\ref{fig:3}, including both
photospheric and external shock contributions. In principle, the neutrino
flavor distribution at the source can be calculated from by
Eqn.\ref{eq:piontonu} and then recomputed at the observer frame after neutrino
oscillations. However the large range of distances, source geometry and density
distribution introduce a large variability in final result, so here simply
approximate the received neutrino flux as having equal numbers in all three
flavors. In this type of models the dominant neutrino emission comes from the
magnetic photosphere. The neutrino spectrum from the external shock peaks at a
higher energy because the magnetic fields and photon densities there are lower
than in the photosphere, and the charged mesons suffer much less synchrotron
cooling before they decay. A uniformly distributed interstellar density of
$n_{\rm ISM}=100~\rm cm^{-3}$ is assumed for the external shock calculation.
The neutrino fluence is nonetheless low compared to the photospheric fluence,
due to the low target photon and proton column density and therefore the low
$pp$ or $p\gamma$ collision rate. A smaller value of $n_{\rm ISM}$ would lead
to a more distant shock, and even lower neutrino fluences.


Fig.\ref{fig:3}, as well as Fig.\ref{fig:4},\ref{fig:5} also illustrate the
effect on the neutrino spectra of including additional physical processes
besides the $p\gamma$ production from the $\Delta^+$-resonance used in many
previous studies, including the recent IceCube GRB data analyses
\citep{Abbasi+12grbnu-nat}. The processes included in Figs.\ref{fig:3},
\ref{fig:4} and the rest are the production of $\pi^{\pm}$ via $\Delta^+$ in
$pp$ as well as $p\gamma$, as well as multi-pion production and $K^{\pm}$ in
$pp$ and $p\gamma$.  We note that $pp$ collisions can become important at lower
energies (before $p\gamma$ interactions set in). This is especially true if the
dissipation radius $R_{\rm dis}$ is small, where the $pp$ collision optical depth
$\tau_{pp}$ approaches unity or above. We also note that the neutrino fluence
is comparable to the photon fluence in fig.\ref{fig:1}. This may be a
relatively conservative value; in some other works a higher acceleration
efficiency , pionization efficiency and proton luminosity e.g
$L_{p}=10L_{\gamma}$ are adopted corresponding to a neutrino luminosity
$L_{\nu}\sim$ a few ${L_{\gamma}}$.


In Fig.\ref{fig:4}, we also show the possible effect of having a dissipative
photosphere at a radius larger than that given by Eqn.(\ref{eq:Rph}), e.g. due
to pair effects. Without specifiyng  an underlying mechanism giving rise to the
dissipation and Band function spectrum this increase is uncertain. If the
photon spectrum is generated by the scattering of thermal electrons associated
with turbulent Alfven waves \citep{Thompson94}, the maximum energy of the
comoving photons can hardly exceed the electron rest mass $m_{e}c^2$ and
negligible pairs are produced. On the other hand if the prompt emission is due
to synchrotron radiation from Fermi-I accelerated electrons e.g.
\citep{Drury+12} and the attenuation beyond $E_{\rm max}^{\prime}$ is due to
$\gamma\gamma \to e^{+}e^{-}$ process, we can calculate $\tau_{\gamma\gamma}$
and the amount of pairs. The existence of pairs with such parameters
(Fig.\ref{fig:1}) could boost the pair-photosphere radius to $\sim20$ times the
original $R_{\rm ph}$.  In the collisional scenario where  protons and neutrons
decouple \citep{Beloborodov10pn}, high energy $\gamma$s and injected poisitrons
from decaying pions induce leptonic cascades and significant pair production,
which gives a similar boost of $\sim 20$ to a pair-photosphere radius.  Without
going into specific model details, we just consider for simplicity the
dissipative and spectrum formation effects associated with such larger
effective pair photospheres, and use this radius for calculating the neutrino
spectrum.  In Fig.\ref{fig:4} the curve labelled ${\rm M},e^{\pm}{\rm Ph}$ is for a
magnetized dynamics photosphere a factor $\sim 20$ larger, and the curve
labelled ${\rm B},e^{\pm}{\rm Ph}$ is for a baryonic photosphere a factor $\sim 3.6$
larger than the value of Eqn.(\ref{eq:Rph}).

The case of an internal shock is also shown, for a dissipation radius estimated
by Eqn.\ref{eq:Ris} and variability time $t_{\rm var}=1$ ms.  Qualitatively,
the harder spectra from larger radius dissipation regions arise because the
magnetic field and photon density is lower at larger radii, reducing the pion
and muon electromagnetic cooling.

The effect of a magnetic versus a baryonic jet bulk dynamics on the neutrino
signatures are compared in Fig.\ref{fig:5}, where we plot the neutrino spectra
from a magnetic photosphere where $\Gamma\propto{r^{1/3}}$, a baryonic
photosphere where $\Gamma\propto{r}$ and an internal shock model. Each case is
calculated for two different terminal Lorentz factors, $\eta=300$ and
$\eta=1000$.

The neutrino fluence in general decreases for increasing $\eta$ in the baryonic
photosphere model. The analytical expression for the photospheric radius is
$R\propto\eta^{-3}$ in this model (Eqn.\ref{eq:Rph}), and having assumed a fixed Band
function in the {\it comoving} frame, the number density of target photons is
$n_{ph}^{\prime}\sim{u_{ph}^{\prime}/E^{\prime}\Delta{E}^{\prime}}\propto
L/R^{2}\eta^{2}\propto\eta^{4}$.  Therefore, the $p\gamma$ inverse cooling
timescale $t_{p\gamma}^{\prime{-1}}\propto{n_{ph}^{\prime}}\propto\eta^{4}$ We
also have $t_{dyn}^{\prime{-1}}\sim\eta{c}/R\sim\eta^{4}\sim{t}_{ad}^{\prime{-1}}$.
The pionization efficiency $f_{\pi}=t_{p\gamma}^{\prime{-1}}/t_{ad}^{\prime{-1}}$ 
has no dependence on $\eta$ in the leading order approximation, or on $R$, 
which is treated as a function of $\eta$ here. But, the synchrotron cooling for 
$\pi^{\pm}$, $\mu^{\pm}$ ($t_{\rm sync}/t_{\rm dec} \propto \eta^{-4}$
for given $\pi,\mu$ energy) diminishes the neutrino flux for larger $\eta$.
\footnote{the proton radiative cooling is also relevant; however, it is
relatively weak due to their heavy mass compared to pion and muons}. 
From Fig.\ref{fig:5} we see that the peak energy of $\eta=1000$ is smaller than 
$\eta=300$ case in the {\it comoving} frame. This indicates strong pion cooling 
taking place. Therefore a higher $\eta$ is associated with a smaller $\nu$ flux.

For the magnetic photosphere, a larger $\eta$ gives smaller $R_{\rm ph}$ and
$\Gamma_{\rm ph}$. Both are advantageous for neutrino production (although a
higher magnetic field lead to stronger cooling of pions and muons, it is less
severe than the baryonic case, e.g. see $B_{5}@R$ column in Table.\ref{tbl:1}).

In Fig.\ref{fig:6}, upper panel, we compare the luminosity dependence of the
single source spectra for a standard $\eta=300$ of an ``optimistic" internal
shock model (with a high dissipation rate and $t_{\rm var}=1,~10,~100$ ms), a
magnetic photosphere model and a baryonic photosphere model.  The optimistic
internal shock model with $t_{\rm var}=1$ ms gives the highest flux at energies
$\lesssim 100$ TeV, while the magnetic photosphere models have a lower flux at
these energies, with their spectrum also peaking towards higher energies. In
Fig.\ref{fig:7} upper panel we show those cases with $\eta=1000$. We see that
the magnetic photosphere case is the least affected by $\eta$.

An interesting possibility in the case of magnetic dissipation regions, e.g.
due to reconnection, is that they may produce an an accelerated proton spectrum
which is harder than the typical Fermi case of $dN/dE\propto{E^{-2}}$. As
discussed by \citep{Drury+12,Bosch-Ramon+12} for an extreme case all the
protons entering the acceleration process are essentially confined by the
magnetic field and the zero escape probability leads to a spectrum
$dN/dE\propto{E^{-1}}$. If this scenario is valid, the bulk of the energy for
the accelerated protons is concentrated in the high energy end of their
injection spectrum. The neutrino energy associated with these protons lies in
the PeV-EeV energy range. This energy is in the sensitivity range of IceCube
and proposed ARIANNA neutrino detector. However this scenario requires
significant magnetic reconnection process where the toroidal magnetic field is
still present. Charged pions and muons at this energy suffer strong synchrotron
cooling which suppresses the neutrino spectrum significantly. Therefore, no
significant neutrino emission is expected from this scenario (since at lower
energies there are too few protons due to the nature of a $dN/dE\propto{E^{-1}}$
spectrum.)


\section{Diffuse neutrino background from various GRB models and implications}
\label{sec:diffuse}

Since the individual source fluxes are very low, except for the unlikely event
of an extremely nearby occurrence, it is useful to consider the cumulative
diffuse neutrino flux from all GRBs in the sky. We calculate this diffuse flux
based on two methods. Method-I uses a GRB luminosity distribution (luminosity
function) and a redshift distribution \citep{Wanderman+10grbsfr}, given,
respectively, by

\begin{equation} \phi(L_{\gamma})\propto \begin{cases}
(L_{\gamma}/L_{*})^{m_{1}} & L_{\rm min}<L_{\gamma}<L_{*} \\
(L_{\gamma}/L_{*})^{m_{2}} & L_{*}<L_{\gamma}<L_{\rm max} \end{cases}
\label{eq:LF} \end{equation}

\begin{equation} R_{\rm GRB}(z)\propto \begin{cases} (1+z)^{n_{1}} & z<z_{1} \\
(1+z)^{n_{2}} & z>z_{1} \end{cases} \label{eq:Zdist} \end{equation} Here
$L_{\gamma}$ is the peak photon luminosity (here mostly in the 0.1-1 MeV range)
and  we have used the following values: $L_{\rm min}=10^{50}$ erg/s,
$L_{*}=10^{52.5}$ erg/s , $L_{\rm max}=10^{54}$ erg/s, $m_{1}=-0.17$ ,
$m_{2}=-1.44$ , $n_{1}=2.07$ , $n_{2}=-1.36$ , $z_{1}=3.1$ which are best fit
values used in \citep{Wanderman+10grbsfr}. 
The differential comoving rate of GRBs at a redshift $z$ is

\begin{equation} R(z)=\frac{R_{\rm GRB}(z)}{(1+z)}\frac{dV}{dz}\label{eq:Rz}
\end{equation} where V(z) is the comoving volume in the standard ${\rm
\Lambda{CDM}}$ cosmology given by parameters in Eqn.\ref{eq:LCDM} and the
factor (1+z) accounts for time dilation effect due to cosmic expansion. The
differential number of GRB per unit redshift is given by

\begin{equation} {\rm d}N(L_{\gamma},z)=\rho_{0}\phi(L_{\gamma})R(z){\rm
d}{\log}L_{\gamma}{\rm d}z \end{equation} where $\rho_{0}=1.3{\rm /yr/Gpc^{3}}$
is the local rate\footnote{if we use the simplest detection criteria for SWIFT
$dN_{ph}/dt\ge0.4{\rm/cm^{2}/s}$, the number of GRBs which meet this criteria
is about $600/yr$ which is a reasonable all sky rate.} of high luminosity GRBs
(not including a separate family of objects called low luminosity GRBs
\citep{Liang+07lowlumgrb}).


\begin{figure} \includegraphics[width=0.99\columnwidth]{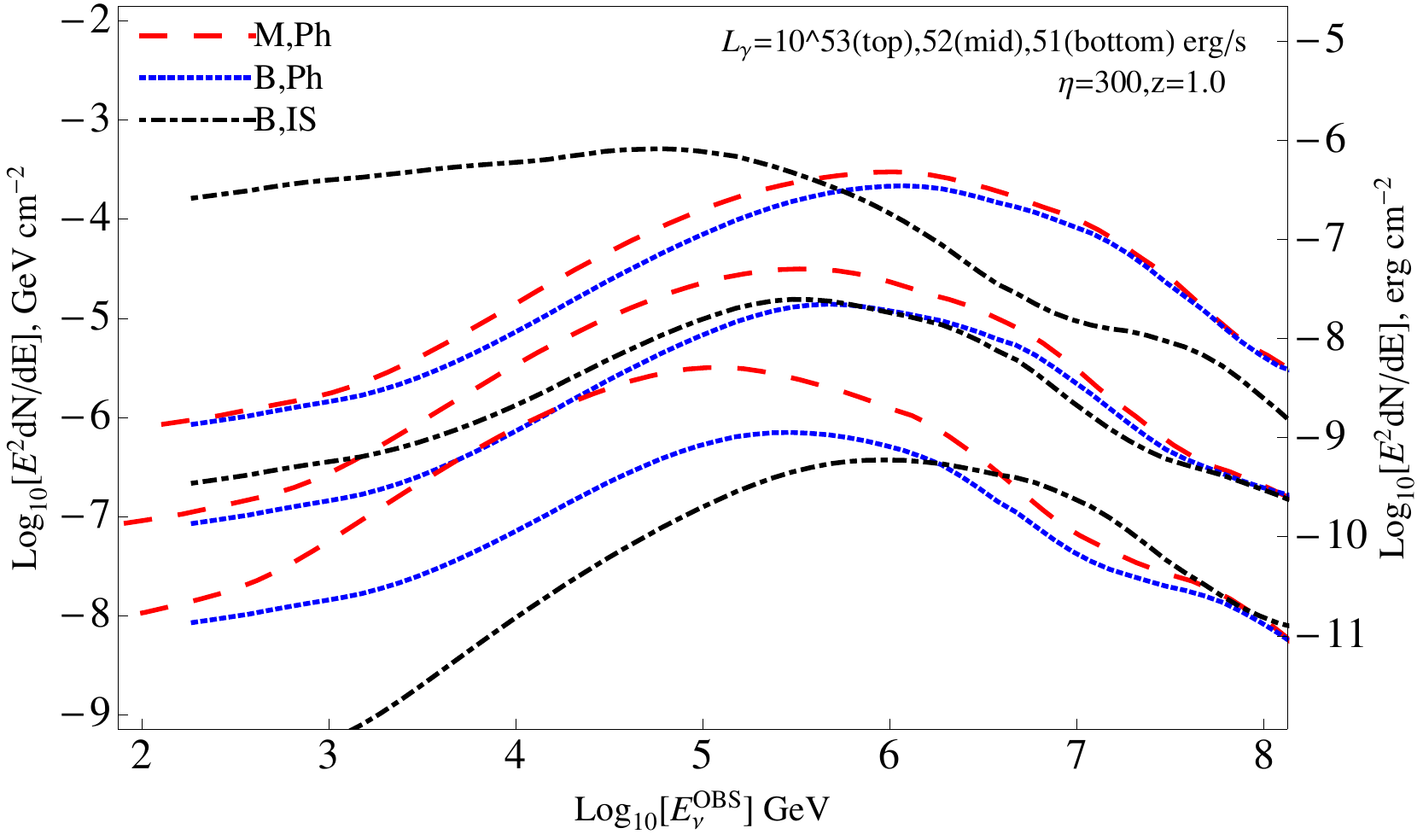}
\includegraphics[width=0.99\columnwidth]{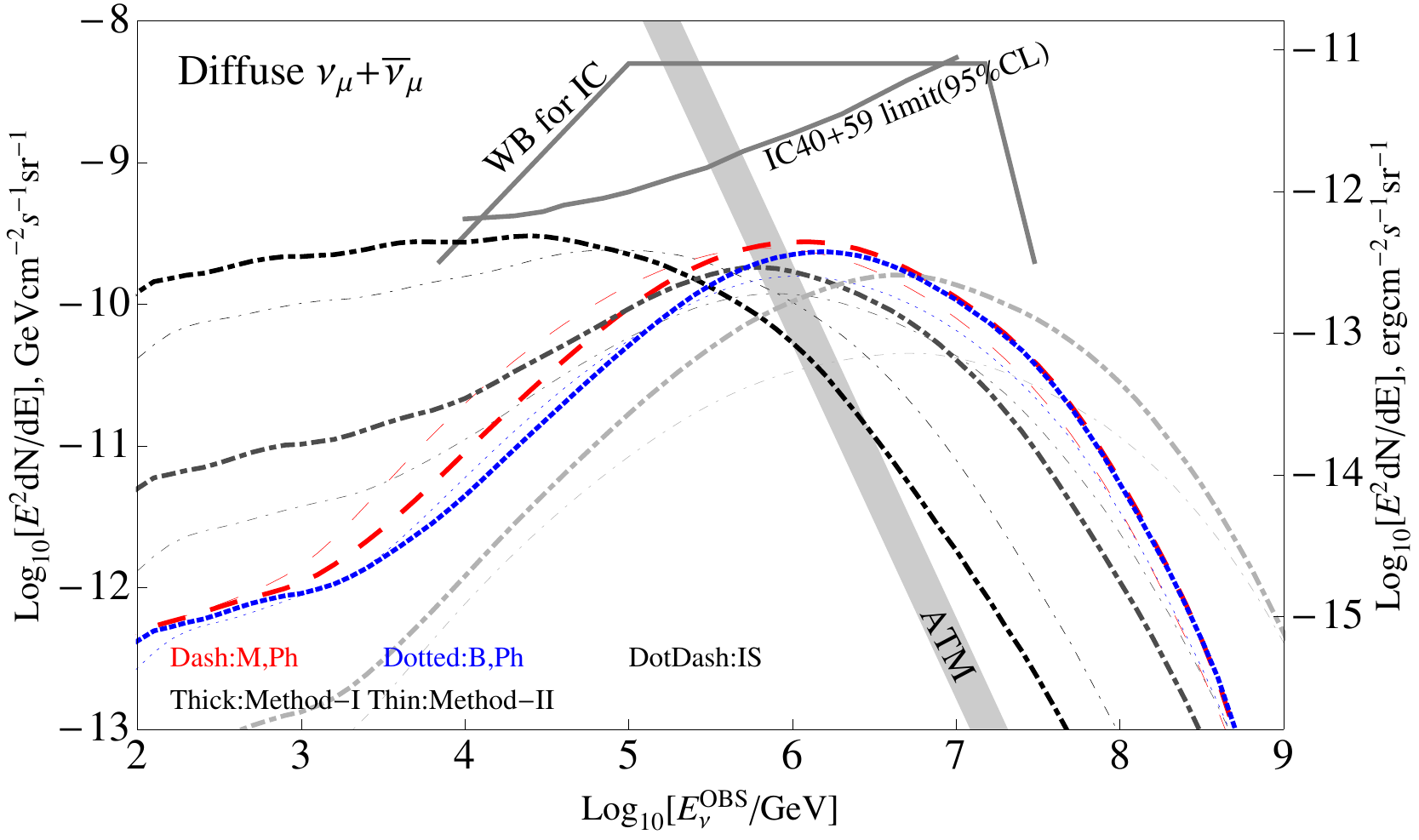} \caption{Upper panel:
Neutrino fluence from a single GRB from different dissipation regions. Red,
dashed: magnetic photosphere; blue, dotted: baryonic photosphere;  Dot-dash:
baryonic internal shock for $t_{\rm var}=1,10,100$ ms. (black, dark gray, light
gray). These are computed for different luminosities (in each model,
$L_{\gamma}=10^{53},10^{52},10^{51}$ erg/s (top,middle,bottom). Lower panel:
Diffuse $\nu_{\mu}+\bar{\nu}_\mu$ neutrino spectral flux from the three models
above (same line style), calculated for an all-sky GRB rate of 700/yr using
statistical Method I (thick lines) and Method II (thin lines; see \S
\ref{sec:diffuse}). Also shown is the IceCube collaboration's representation of
the diffuse flux from a standard Waxman-Bahcall internal shock model, and the
IC 40+59 observational upper limit (see Fig.3 of \citep{Abbasi+12grbnu-nat} for
description). The gray zone labeled ATM is the atmospheric neutrino spectrum.
The plots from both panels suggest that the occasional electromagnetically
bright GRBs can contribute significantly to the total diffuse flux.}
\label{fig:6} \end{figure}


\begin{figure} \includegraphics[width=0.99\columnwidth]{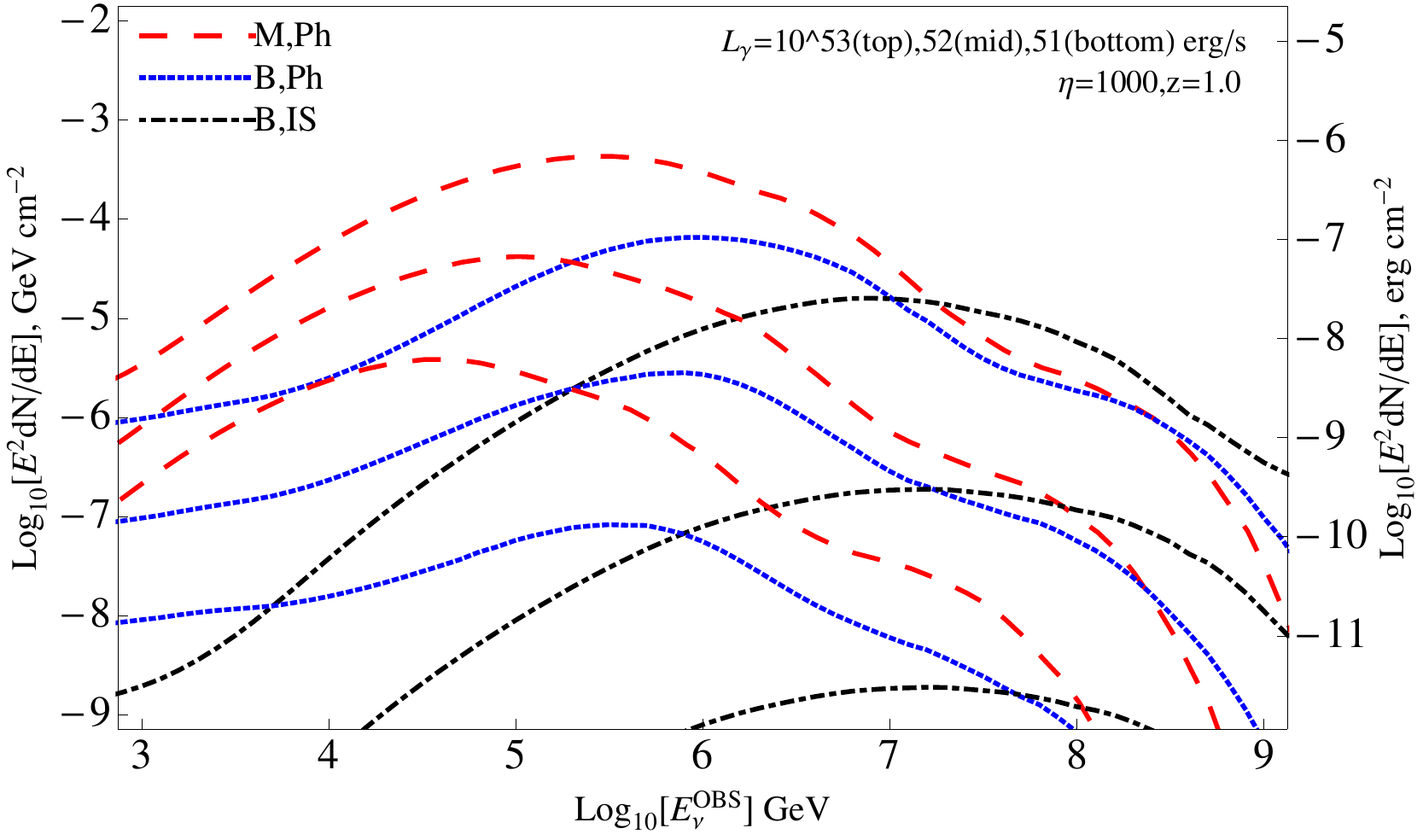}
\includegraphics[width=0.99\columnwidth]{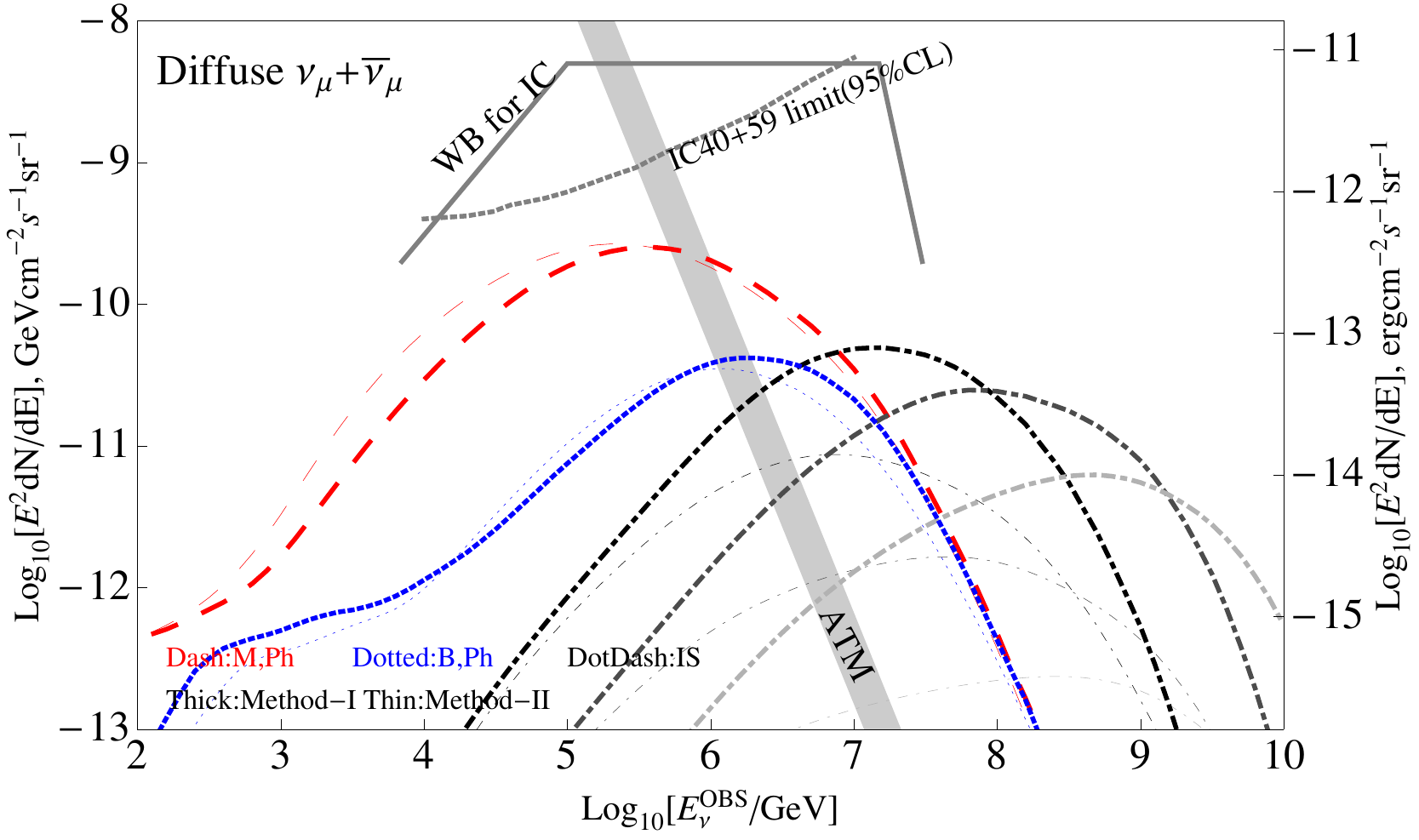} \caption{The parameters and
conventions are the same as those in fig.\ref{fig:6} except we have used
$\eta=1000$ (in fig.\ref{fig:6},$\eta=300$ is used).  } \label{fig:7}
\end{figure}

An alternative method (Method-II) is to use the Wisconsin GRB catalog
[http://grbweb.icecube.wisc.edu/]. We consider only long GRBs and use the
parameters such as the Band photon index, fluence, redshift, photon peak energy
etc. to compute the neutrino spectrum from each individual burst and stack them
together.  Finally we normalize the resultant flux to an all sky rate of 700
GRBs/yr\footnote{Since these instruments have a limited sky coverage and
operation time bin, a smaller number of GRBs per year are actually recorded}.
These two methods are then used to calculate the diffuse neutrino flux assuming
that the GRB neutrinos are due to a magnetic photosphere, a baryonic
photosphere, and an internal shock model (Fig.\ref{fig:6},\ref{fig:7}). In
Fig.\ref{fig:6} we plot those cases with $\eta=300$ while in Fig.\ref{fig:7}
$\eta=1000$.

An inspection of this Fig.\ref{fig:6},\ref{fig:7} (lower panel) shows that the
above models, with the parameters used in this paper, predict a diffuse
neutrino flux which is likely to be within the current constraints set by
IceCube 40+59 string observations, as suggested by the fact that they lie below
the two IceCube constraint lines labeled WB for IC and IC 40+59. A caveat is
that these constraints are upper limits on the flux as a function of the break
energy assuming a Band function with $\alpha=-1,\beta=-2$, or assuming a slope
-2.  They are not directly applicable to other types of spectra; thus, these
constraint lines are here intended only for a rough comparison; they would have
to be re-evaluated for the spectra shown here. Nonetheless, they do provide
some guidance, and it appears that the only models which are close to being
constrained at present  are those using the most optimistic parameters; e.g.
for internal shock models with a $t_{\rm var}=1$ ms one would expect a small
dissipation radius leading to a high $\tau_{\rm pp}$ and a high neutrino
fluence in the lower energy range from $pp$ collisions (where $\tau_{\rm
p\gamma}$ has saturated to unity with $\tau_{\rm pp}$ increasing, making $pp$
collisions more important.) Also, if we were to assume a higher value of
$\epsilon_{p}/\epsilon_{e}$, such as 10 (in this paper we used a value of $1$,
see also section. \ref{sec:single}) and/or if we were to adopt a lower magnetic
field fraction ($\epsilon_{B}\simeq10^{-2}$, rather than the
$\epsilon_{d}\epsilon_B=0.1$ used here), the diffuse neutrino fluence would be
likely to violate the above constraints, especially for the internal shock
model. However, this is for the optimistic internal shock case where one uses
$t_{\rm var}=1$ ms, whereas there is a larger uncertainty in this quantity, and
in most bursts $t_{\rm var}$ can often be several orders or magnitude larger
(see comments in the third paragraph of \S \ref{sec:single}). For such larger
(and more reasonable) values of $t_{\rm var}$ the model would appear to be
still compatible with the constraints, even if a high
$\epsilon_{p}/\epsilon_{e}$ ratio were assumed.

We have also included the approximate atmospheric neutrino background in
Fig.\ref{fig:6},\ref{fig:7}. It is worth noting that GRBs are transient sources
whose prompt emission duration in the observer frame is within $t_{\rm
dur}^{\rm obs}\sim100$ s. The angular resolution for TeV neutrino and above is
within $1~{\rm deg^{2}}$. Therefore, considering the search time bin and the
small solid angle set by optical observations, the effective atmospheric
neutrino background is well below these GRB diffuse fluxes. In other words,
even one or two muon events in IceCube correlated with photon detections would
give a high signal to noise ratio.

A caveat for such calculations of the diffuse background is that the
statistical description of both the source physics parameters and the
spatial-temporal distribution of the sources has large uncertainties; this
applies to both the luminosity function (Method I) and the observed burst
catalog (Method II). In Method-I, the parameters in eqns. (\ref{eq:LF},
\ref{eq:Zdist}) have large uncertainties; especially in the low redshift region
(e.g. $z<0.3$), very few GRBs are observed. However, in order to expect more
than one muon event in IceCube, we need a GRB of moderate or high luminosity
located at low redshift (such as the example GRB in fig.\ref{fig:4} with
$L_{\gamma}=10^{52.5}$ erg/s and $z=0.1$, which results in about 0.2 muon
events in the IceCube 86-string configuration.) The estimated rate of GRBs
which satisfy $N_{\rm \mu,Icecube}\ge{1}$  is about $o(1)$ per ten years. In
Method-II, we actually have a very limited number of GRBs with well measured
redshifts in the catalog. Those without a redshift are assigned a default
redshift value of $z=2.0$.  Thus, improvements in the photon-based statistics
of bursts, as well as neutrino observations over multi-year periods appear
required.


\section{Discussion} \label{sec:disc}

We have calculated the $\nu_\mu + {\bar \nu}_\mu$ signatures expected from
photospheric GRB models where the dynamics is either magnetically or
baryonically dominated, including also the effects of external shocks, and have
compared these signatures with those expected from the baryonic internal shock
models. This comparison is timely in view of recent developments, the most
pressing being the recently published IceCube constraints
\citep{Abbasi+11-ic40diff,Abbasi+12grbnu-nat} on the ``standard" internal shock
GRB models. The exploration of alternative models to the internal shocks has
its own separate motivation, independently of the IceCube observations. One
reason is the increased realization that magnetic fields may play a dominant
role in the GRB phenomenon, and the dynamics of magnetically dominated GRB jet
models \footnote{Magnetically dominated jets are naturally expected if the
black hole energy is extracted via a Blandford-Znajek mechanism
\citep{Blandford+77znajek} , or if the source is a temporary magnetar
\citep{Metzger+11grbmag}.} are being considered in detail \citep{Drenkhahn+02,
Tchekhovskoy+10grb,McKinney+12magphot}.  Also, issues related to the efficiency
and spectrum of standard internal shock models of the prompt $\gamma$-ray
emission \citep{Preece+00batse, Medvedev+04magshock, Meszaros06grbrev} have
led, on the one hand, to considering modified internal shock models that
address these issues, e.g.\citep{Asano+11grbtemp,Inoue+11shockturb,
Murase+12reac}, etc., and, on the other hand, to considering the photospheric
emission as the source of the prompt $\gamma$-rays, including both baryonic
photospheres \citep{Meszaros+00phot, Ryde05,Rees+05phot, Peer+06phot,
Beloborodov10pn, Peer11-fermigrb} and magnetic photospheres
\citep{Giannios+07photspec,Meszaros+11col,Veres+12mag}.  Some previous
calculations of neutrino spectra from baryonically dominated photospheres have
been carried out \citep{Murase+08photonu,Wang+09photonu}, but so far none from
dissipative photospheres\footnote{ As we were ready to submit we received a 
preprint on this subject \citep{Zhang+12grbnu}, with results compatible with ours.}
or magnetic photospheres, which we treat here

One of the notable features about the neutrino emission from dissipative
photosphere models, both in magnetized of baryonic dynamics, is that the peak
energy of the spectrum (which is around 0.5-1 PeV, as seen in Fig. \ref{fig:3})
differs from the peak energies of a typical Waxman-Bahcall
(WB97)\citep{Waxman+97grbnu} simplified standard internal shock model such as
used in the recent IceCube studies (and in
\citep{Guetta+01grbnu,Guetta+04grbnu}). The peak energy is also higher than
those for the internal shock model with $t_{\rm var}=1{\rm ms}$ and $\eta=300$,
but lower than those with higher $t_{\rm var}$ and $\eta$ values. In part, this
is due to the inclusion, in addition to $\Delta$ resonance production, also of
multi-pion effects, Kaon decay, etc., which also gives a steeper spectrum above
the peak, compared to the flat slope of $E^{2} dN/dE$ above the peak expected
in the usual WB97 internal shock model.  This steepening of the spectrum above
the peak is not unique to the photospheric dissipation model, it occurs also in
internal shocks when we include the above additional physics beyond the
$\Delta$-resonance, as can be seen in Fig. \ref{fig:4}. (Such a steepening for
internal shocks is also found by \citep{Hummer+11nu-ic3}).

The energy at which the spectral peak appears depends generically on the radius
of the dissipative and photon escape zones, which here we have assumed to be
colocated, whether it be a photosphere or an internal shock. The effect of
different radii is illustrated in Fig. \ref{fig:4}, which  compares the spectra
and peak energies of different photospheric dissipation zones at different
radii and also two internal shocks at two different radii (or variability
times).  The harder spectra from larger radii are expected because of the
decrease of the magnetic fields and and photon densities, which allows the
higher energy secondary pions and muons to decay before significant
electromagnetic cooling has taken place.

The difference between a magnetically dominated photosphere (M,Ph), a
baryonically dominated photosphere (B,Ph) and internal shock (IS) models are
compared in Fig.\ref{fig:4},\ref{fig:5}, Fig.\ref{fig:6},\ref{fig:7}(upper
panel) and Table.\ref{tbl:1}. The dissipation region for the `M,Ph' typically
lies in the acceleration phase of the jet where there is a smaller $\Gamma_{\rm
ph}$ than `B,Ph' and `IS' model. The neutrino spectrum in the observer frame is
affected by the efficiency of $pp$ and $p\gamma$ interaction, the cooling of
the secondary charged particles (which is mainly synchrotron, determined by the
comoving magnetic field) and finally Lorentz boost and redshift. The
differences in neutrino spectra from these models are not quite significant for
$\eta=300$ where the dissipation region, magnetic field and Lorentz factor are
similar. For $\eta=1000$, the `B,Ph' and `IS' have harder spectra than the
`M,Ph' model. The `B,Ph' radius here is rather small where $pp$, $p\gamma$ and
coolings are all efficient. The large Lorentz boost factor finally pushes the
peak neutrino energy in the observer frame over the `M,Ph' model. For `IS'
model with $\eta=1000$ the dissipation radii is much larger than `M,Ph' and
`B,Ph' case. The inefficient cooling results in the highest peak energy of the
three models. However, the flux is also the lowest because $pp$ and $p\gamma$
are the least efficient of these models. 

The maximum source-frame energy of the accelerated protons which are able to
escape the acceleration region, for the models considered here, are shown in
the third from the last column of Table.\ref{tbl:1}. Leaving out any consideration of the
diffuse flux, it is seen that only the internal shock models approach the
highest energies associated with the GZK limit, as originally suggested by
\citep{Waxman+97grbnu}, while all photospheric models fall several orders of
magnitude below this energy. We have not done an exhaustive parameter
search\footnote{For standard (unmodified) internal shock models, such searches
have been done by e.g.\citep{Guetta+01is,Kotera+11uhecr, Ahlers+11-grbprob}},
since our emphasis has been the neutrino emission, but it is apparent that
photospheric models would not be competitive GZK sources.

The detection of neutrinos from individual single sources with the 86 string
IceCube, as seen from Table.\ref{tbl:1} in the last two columns, is rather
difficult, the number of expected muon events at best being $\sim 0.2$ for a
magnetic photosphere model of average luminosity at $z=0.1$. The only hope may be 
the statistically rare observation of a nearby bright GRB , or through observations 
of the diffuse flux.

The diffuse flux offers higher prospects for an eventual detection, and a
comparison of the diffuse flux expected from the three different types of
models discussed, Figs. \ref{fig:6} and \ref{fig:7} lower panels, shows that an
internal shock scenario with optimistic parameters such as $t_{\rm var}=1{\rm
ms}$ and $\eta=300$ comes closest to being constrained by the current IceCube
limits \citep{Abbasi+12grbnu-nat}.  This confirms the recent calculations of
\citep{Li11-nu-ic3,He+12grbnu,Hummer+11nu-ic3}.  On the other hand, a magnetic
photosphere model is far from being constrained by the current limits (although
the limits will have to be re-evaluated for the specific spectral shapes).  A
baryonic dissipative photosphere model has an even lower flux, and would take
the longest to be detected. Fig.\ref{fig:7} shows that a higher Lorentz factor,
in this case $\eta=1000$, decreases the constraints. The diffuse flux from a
magnetic photosphere is the least affected, and both baryonic photosphere and
internal shock models have higher peak energy but lower flux.

In conclusion, the calculations discussed in the previous sections indicate
that the current IceCube limits are not yet sufficient to distinguish between
the dissipative photosheric models based on magnetic or baryonic dynamics,
although they are approaching meaningful limits for baryonic internal shocks
outside the photospheres. Also, as of now it is not yet possible to set
stringent limits on the value of $\epsilon_{p}/\epsilon_{e}$, i.e. the putative
ration of accelerated protons to accelerated electrons.  One of the reasons,
for the photospheric models, is that the magnetic reconnection or dissipation
scenario is less straightforward than the simple internal shock scenario,
having both more physical model uncertainties and parameters associated with
them, e.g. the geometry of the plasma layers or the striped field structures,
resistivity, instabilities, reconnection rate, etc.  Both for photospheric and
internal shock models, the fraction of protons that are injected into the
reconnection or Fermi acceleration process is uncertain. The acceleration
timescale and escape probability depend on the geometry and dynamics of
magnetic reconnection or acceleration region. These factors are also crucial
for determining the injection of a cosmic ray proton spectrum and its effect on
the neutrino fluence, maximum neutrino energy and spectrum.

For these reasons, and in preparation for future improved limits from longer
observation times and from possible future higher sensitivity detectors, we
have here investigated the typical neutrino spectral features and flux levels
expected from three of the basic GRB models currently being considered.  There
are reasonable prospects that the detection or non-detection of these neutrino
fluxes in the next decade with IceCube or next generation of large neutrino
detectors will shed light on the underlying physics of the GRB jets and on the
cosmic ray acceleration process in then.

\acknowledgments{We are indebted to Peter Veres, Binbin Zhang, Kazumi Kashiyama
and Kohta Murase for discussions, Bing Zhang and Pawan Kumar for communications,
and to NSF PHY-0757155 (SG and PM) and Grant-in-Aid for Scientific Research 
No.22740117 from the Ministry of Education, Culture, Sports, Science and Technology
(MEXT) of Japan (KA) for partial financial support.}


\begin{thebibliography}{61}%
\makeatletter
\providecommand \@ifxundefined [1]{%
 \@ifx{#1\undefined}
}%
\providecommand \@ifnum [1]{%
 \ifnum #1\expandafter \@firstoftwo
 \else \expandafter \@secondoftwo
 \fi
}%
\providecommand \@ifx [1]{%
 \ifx #1\expandafter \@firstoftwo
 \else \expandafter \@secondoftwo
 \fi
}%
\providecommand \natexlab [1]{#1}%
\providecommand \enquote  [1]{``#1''}%
\providecommand \bibnamefont  [1]{#1}%
\providecommand \bibfnamefont [1]{#1}%
\providecommand \citenamefont [1]{#1}%
\providecommand \href@noop [0]{\@secondoftwo}%
\providecommand \href [0]{\begingroup \@sanitize@url \@href}%
\providecommand \@href[1]{\@@startlink{#1}\@@href}%
\providecommand \@@href[1]{\endgroup#1\@@endlink}%
\providecommand \@sanitize@url [0]{\catcode `\\12\catcode `\$12\catcode
  `\&12\catcode `\#12\catcode `\^12\catcode `\_12\catcode `\%12\relax}%
\providecommand \@@startlink[1]{}%
\providecommand \@@endlink[0]{}%
\providecommand \url  [0]{\begingroup\@sanitize@url \@url }%
\providecommand \@url [1]{\endgroup\@href {#1}{\urlprefix }}%
\providecommand \urlprefix  [0]{URL }%
\providecommand \Eprint [0]{\href }%
\providecommand \doibase [0]{http://dx.doi.org/}%
\providecommand \selectlanguage [0]{\@gobble}%
\providecommand \bibinfo  [0]{\@secondoftwo}%
\providecommand \bibfield  [0]{\@secondoftwo}%
\providecommand \translation [1]{[#1]}%
\providecommand \BibitemOpen [0]{}%
\providecommand \bibitemStop [0]{}%
\providecommand \bibitemNoStop [0]{.\EOS\space}%
\providecommand \EOS [0]{\spacefactor3000\relax}%
\providecommand \BibitemShut  [1]{\csname bibitem#1\endcsname}%
\let\auto@bib@innerbib\@empty
\bibitem [{\citenamefont {Waxman}\ and\ \citenamefont
  {Bahcall}(1997)}]{Waxman+97grbnu}%
  \BibitemOpen
  \bibfield  {author} {\bibinfo {author} {\bibfnamefont {E.}~\bibnamefont
  {Waxman}}\ and\ \bibinfo {author} {\bibfnamefont {J.}~\bibnamefont
  {Bahcall}},\ }\href@noop {} {\bibfield  {journal} {\bibinfo  {journal}
  {{\prl}}\ }\textbf {\bibinfo {volume} {78}},\ \bibinfo {pages} {2292}
  (\bibinfo {year} {1997})}\BibitemShut {NoStop}%
\bibitem [{\citenamefont {{Guetta}}\ \emph {et~al.}(2004)\citenamefont
  {{Guetta}}, \citenamefont {{Hooper}}, \citenamefont {{Alvarez-Mu{\~n}iz}},
  \citenamefont {{Halzen}},\ and\ \citenamefont {{Reuveni}}}]{Guetta+04grbnu}%
  \BibitemOpen
  \bibfield  {author} {\bibinfo {author} {\bibfnamefont {D.}~\bibnamefont
  {{Guetta}}}, \bibinfo {author} {\bibfnamefont {D.}~\bibnamefont {{Hooper}}},
  \bibinfo {author} {\bibfnamefont {J.}~\bibnamefont {{Alvarez-Mu{\~n}iz}}},
  \bibinfo {author} {\bibfnamefont {F.}~\bibnamefont {{Halzen}}}, \ and\
  \bibinfo {author} {\bibfnamefont {E.}~\bibnamefont {{Reuveni}}},\ }\href
  {\doibase 10.1016/S0927-6505(03)00211-1} {\bibfield  {journal} {\bibinfo
  {journal} {Astroparticle Physics}\ }\textbf {\bibinfo {volume} {20}},\
  \bibinfo {pages} {429} (\bibinfo {year} {2004})},\ \Eprint
  {http://arxiv.org/abs/arXiv:astro-ph/0302524} {arXiv:astro-ph/0302524}
  \BibitemShut {NoStop}%
\bibitem [{\citenamefont {{Abbasi}}\ \emph {et~al.}(2011)\citenamefont
  {{Abbasi}}, \citenamefont {{Abdou}}, \citenamefont {{Abu-Zayyad}},
  \citenamefont {{Adams}}, \citenamefont {{Aguilar}}, \citenamefont {{Ahlers}},
  \citenamefont {{Altmann}}, \citenamefont {{Andeen}}, \citenamefont
  {{Auffenberg}}, \citenamefont {{Bai}},\ and\ \citenamefont
  {et~al.}}]{Abbasi+11-ic40diff}%
  \BibitemOpen
  \bibfield  {author} {\bibinfo {author} {\bibfnamefont {R.}~\bibnamefont
  {{Abbasi}}}, \bibinfo {author} {\bibfnamefont {Y.}~\bibnamefont {{Abdou}}},
  \bibinfo {author} {\bibfnamefont {T.}~\bibnamefont {{Abu-Zayyad}}}, \bibinfo
  {author} {\bibfnamefont {J.}~\bibnamefont {{Adams}}}, \bibinfo {author}
  {\bibfnamefont {J.~A.}\ \bibnamefont {{Aguilar}}}, \bibinfo {author}
  {\bibfnamefont {M.}~\bibnamefont {{Ahlers}}}, \bibinfo {author}
  {\bibfnamefont {D.}~\bibnamefont {{Altmann}}}, \bibinfo {author}
  {\bibfnamefont {K.}~\bibnamefont {{Andeen}}}, \bibinfo {author}
  {\bibfnamefont {J.}~\bibnamefont {{Auffenberg}}}, \bibinfo {author}
  {\bibfnamefont {X.}~\bibnamefont {{Bai}}}, \ and\ \bibinfo {author}
  {\bibnamefont {et~al.}},\ }\href {\doibase 10.1103/PhysRevD.84.082001}
  {\bibfield  {journal} {\bibinfo  {journal} {\prd}\ }\textbf {\bibinfo
  {volume} {84}},\ \bibinfo {eid} {082001} (\bibinfo {year} {2011})},\ \Eprint
  {http://arxiv.org/abs/1104.5187} {arXiv:1104.5187 [astro-ph.HE]} \BibitemShut
  {NoStop}%
\bibitem [{\citenamefont {{Abbasi}}\ \emph {et~al.}(2012)\citenamefont
  {{Abbasi}}, \citenamefont {{Abdou}}, \citenamefont {{Abu-Zayyad}},
  \citenamefont {{Ackermann}}, \citenamefont {{Adams}}, \citenamefont
  {{Aguilar}}, \citenamefont {{Ahlers}}, \citenamefont {{Altmann}},
  \citenamefont {{Andeen}}, \citenamefont {{Auffenberg}},\ and\ \citenamefont
  {et~al.}}]{Abbasi+12grbnu-nat}%
  \BibitemOpen
  \bibfield  {author} {\bibinfo {author} {\bibfnamefont {R.}~\bibnamefont
  {{Abbasi}}}, \bibinfo {author} {\bibfnamefont {Y.}~\bibnamefont {{Abdou}}},
  \bibinfo {author} {\bibfnamefont {T.}~\bibnamefont {{Abu-Zayyad}}}, \bibinfo
  {author} {\bibfnamefont {M.}~\bibnamefont {{Ackermann}}}, \bibinfo {author}
  {\bibfnamefont {J.}~\bibnamefont {{Adams}}}, \bibinfo {author} {\bibfnamefont
  {J.~A.}\ \bibnamefont {{Aguilar}}}, \bibinfo {author} {\bibfnamefont
  {M.}~\bibnamefont {{Ahlers}}}, \bibinfo {author} {\bibfnamefont
  {D.}~\bibnamefont {{Altmann}}}, \bibinfo {author} {\bibfnamefont
  {K.}~\bibnamefont {{Andeen}}}, \bibinfo {author} {\bibfnamefont
  {J.}~\bibnamefont {{Auffenberg}}}, \ and\ \bibinfo {author} {\bibnamefont
  {et~al.}},\ }\href {\doibase 10.1038/nature11068} {\bibfield  {journal}
  {\bibinfo  {journal} {\nat}\ }\textbf {\bibinfo {volume} {484}},\ \bibinfo
  {pages} {351} (\bibinfo {year} {2012})},\ \Eprint
  {http://arxiv.org/abs/1204.4219} {arXiv:1204.4219 [astro-ph.HE]} \BibitemShut
  {NoStop}%
\bibitem [{\citenamefont {{Ghisellini}}\ and\ \citenamefont
  {{Celotti}}(1999)}]{Ghisellini+99grbspec}%
  \BibitemOpen
  \bibfield  {author} {\bibinfo {author} {\bibfnamefont {G.}~\bibnamefont
  {{Ghisellini}}}\ and\ \bibinfo {author} {\bibfnamefont {A.}~\bibnamefont
  {{Celotti}}},\ }\href {\doibase 10.1051/aas:1999339} {\bibfield  {journal}
  {\bibinfo  {journal} {\aaps}\ }\textbf {\bibinfo {volume} {138}},\ \bibinfo
  {pages} {527} (\bibinfo {year} {1999})},\ \Eprint
  {http://arxiv.org/abs/arXiv:astro-ph/9906145} {arXiv:astro-ph/9906145}
  \BibitemShut {NoStop}%
\bibitem [{\citenamefont {{Preece}}\ \emph {et~al.}(2000)\citenamefont
  {{Preece}}, \citenamefont {{Briggs}}, \citenamefont {{Mallozzi}},
  \citenamefont {{Pendleton}}, \citenamefont {{Paciesas}},\ and\ \citenamefont
  {{Band}}}]{Preece+00batse}%
  \BibitemOpen
  \bibfield  {author} {\bibinfo {author} {\bibfnamefont {R.~D.}\ \bibnamefont
  {{Preece}}}, \bibinfo {author} {\bibfnamefont {M.~S.}\ \bibnamefont
  {{Briggs}}}, \bibinfo {author} {\bibfnamefont {R.~S.}\ \bibnamefont
  {{Mallozzi}}}, \bibinfo {author} {\bibfnamefont {G.~N.}\ \bibnamefont
  {{Pendleton}}}, \bibinfo {author} {\bibfnamefont {W.~S.}\ \bibnamefont
  {{Paciesas}}}, \ and\ \bibinfo {author} {\bibfnamefont {D.~L.}\ \bibnamefont
  {{Band}}},\ }\href {\doibase 10.1086/313289} {\bibfield  {journal} {\bibinfo
  {journal} {\apjs}\ }\textbf {\bibinfo {volume} {126}},\ \bibinfo {pages} {19}
  (\bibinfo {year} {2000})},\ \Eprint
  {http://arxiv.org/abs/arXiv:astro-ph/9908119} {arXiv:astro-ph/9908119}
  \BibitemShut {NoStop}%
\bibitem [{\citenamefont {{Medvedev}}\ \emph {et~al.}(2004)\citenamefont
  {{Medvedev}}, \citenamefont {{Silva}}, \citenamefont {{Fiore}}, \citenamefont
  {{Fonseca}},\ and\ \citenamefont {{Mori}}}]{Medvedev+04magshock}%
  \BibitemOpen
  \bibfield  {author} {\bibinfo {author} {\bibfnamefont {M.~V.}\ \bibnamefont
  {{Medvedev}}}, \bibinfo {author} {\bibfnamefont {L.~O.}\ \bibnamefont
  {{Silva}}}, \bibinfo {author} {\bibfnamefont {M.}~\bibnamefont {{Fiore}}},
  \bibinfo {author} {\bibfnamefont {R.~A.}\ \bibnamefont {{Fonseca}}}, \ and\
  \bibinfo {author} {\bibfnamefont {W.~B.}\ \bibnamefont {{Mori}}},\
  }\href@noop {} {\bibfield  {journal} {\bibinfo  {journal} {Journal of Korean
  Astronomical Society}\ }\textbf {\bibinfo {volume} {37}},\ \bibinfo {pages}
  {533} (\bibinfo {year} {2004})}\BibitemShut {NoStop}%
\bibitem [{\citenamefont {Meszaros}(2006)}]{Meszaros06grbrev}%
  \BibitemOpen
  \bibfield  {author} {\bibinfo {author} {\bibfnamefont {P.}~\bibnamefont
  {Meszaros}},\ }\href@noop {} {\bibfield  {journal} {\bibinfo  {journal}
  {Rept. Prog. Phys.}\ }\textbf {\bibinfo {volume} {69}},\ \bibinfo {pages}
  {2259} (\bibinfo {year} {2006})},\ \Eprint
  {http://arxiv.org/abs/astro-ph/0605208} {astro-ph/0605208} \BibitemShut
  {NoStop}%
\bibitem [{\citenamefont {{Asano}}\ and\ \citenamefont
  {{Terasawa}}(2009)}]{Asano+09slowheat}%
  \BibitemOpen
  \bibfield  {author} {\bibinfo {author} {\bibfnamefont {K.}~\bibnamefont
  {{Asano}}}\ and\ \bibinfo {author} {\bibfnamefont {T.}~\bibnamefont
  {{Terasawa}}},\ }\href {\doibase 10.1088/0004-637X/705/2/1714} {\bibfield
  {journal} {\bibinfo  {journal} {\apj}\ }\textbf {\bibinfo {volume} {705}},\
  \bibinfo {pages} {1714} (\bibinfo {year} {2009})},\ \Eprint
  {http://arxiv.org/abs/0905.1392} {arXiv:0905.1392 [astro-ph.HE]} \BibitemShut
  {NoStop}%
\bibitem [{\citenamefont {{Inoue}}\ \emph {et~al.}(2011)\citenamefont
  {{Inoue}}, \citenamefont {{Asano}},\ and\ \citenamefont
  {{Ioka}}}]{Inoue+11shockturb}%
  \BibitemOpen
  \bibfield  {author} {\bibinfo {author} {\bibfnamefont {T.}~\bibnamefont
  {{Inoue}}}, \bibinfo {author} {\bibfnamefont {K.}~\bibnamefont {{Asano}}}, \
  and\ \bibinfo {author} {\bibfnamefont {K.}~\bibnamefont {{Ioka}}},\ }\href
  {\doibase 10.1088/0004-637X/734/2/77} {\bibfield  {journal} {\bibinfo
  {journal} {\apj}\ }\textbf {\bibinfo {volume} {734}},\ \bibinfo {eid} {77}
  (\bibinfo {year} {2011})},\ \Eprint {http://arxiv.org/abs/1011.6350}
  {arXiv:1011.6350 [astro-ph.HE]} \BibitemShut {NoStop}%
\bibitem [{\citenamefont {{Murase}}\ \emph {et~al.}(2012)\citenamefont
  {{Murase}}, \citenamefont {{Asano}}, \citenamefont {{Terasawa}},\ and\
  \citenamefont {{M{\'e}sz{\'a}ros}}}]{Murase+12reac}%
  \BibitemOpen
  \bibfield  {author} {\bibinfo {author} {\bibfnamefont {K.}~\bibnamefont
  {{Murase}}}, \bibinfo {author} {\bibfnamefont {K.}~\bibnamefont {{Asano}}},
  \bibinfo {author} {\bibfnamefont {T.}~\bibnamefont {{Terasawa}}}, \ and\
  \bibinfo {author} {\bibfnamefont {P.}~\bibnamefont {{M{\'e}sz{\'a}ros}}},\
  }\href {\doibase 10.1088/0004-637X/746/2/164} {\bibfield  {journal} {\bibinfo
   {journal} {\apj}\ }\textbf {\bibinfo {volume} {746}},\ \bibinfo {eid} {164}
  (\bibinfo {year} {2012})},\ \Eprint {http://arxiv.org/abs/1107.5575}
  {arXiv:1107.5575 [astro-ph.HE]} \BibitemShut {NoStop}%
\bibitem [{\citenamefont {{Zhang}}\ and\ \citenamefont
  {{Yan}}(2011)}]{ZhangICMART}%
  \BibitemOpen
  \bibfield  {author} {\bibinfo {author} {\bibfnamefont {B.}~\bibnamefont
  {{Zhang}}}\ and\ \bibinfo {author} {\bibfnamefont {H.}~\bibnamefont
  {{Yan}}},\ }\href {\doibase 10.1088/0004-637X/726/2/90} {\bibfield  {journal}
  {\bibinfo  {journal} {\apj}\ }\textbf {\bibinfo {volume} {726}},\ \bibinfo
  {eid} {90} (\bibinfo {year} {2011})},\ \Eprint
  {http://arxiv.org/abs/1011.1197} {arXiv:1011.1197 [astro-ph.HE]} \BibitemShut
  {NoStop}%
\bibitem [{\citenamefont {{M{\'e}sz{\'a}ros}}\ and\ \citenamefont
  {{Rees}}(2000)}]{Meszaros+00phot}%
  \BibitemOpen
  \bibfield  {author} {\bibinfo {author} {\bibfnamefont {P.}~\bibnamefont
  {{M{\'e}sz{\'a}ros}}}\ and\ \bibinfo {author} {\bibfnamefont {M.~J.}\
  \bibnamefont {{Rees}}},\ }\href {\doibase 10.1086/308371} {\bibfield
  {journal} {\bibinfo  {journal} {\apj}\ }\textbf {\bibinfo {volume} {530}},\
  \bibinfo {pages} {292} (\bibinfo {year} {2000})},\ \Eprint
  {http://arxiv.org/abs/arXiv:astro-ph/9908126} {arXiv:astro-ph/9908126}
  \BibitemShut {NoStop}%
\bibitem [{\citenamefont {{Ryde}}(2005)}]{Ryde05}%
  \BibitemOpen
  \bibfield  {author} {\bibinfo {author} {\bibfnamefont {F.}~\bibnamefont
  {{Ryde}}},\ }\href {\doibase 10.1086/431239} {\bibfield  {journal} {\bibinfo
  {journal} {\apjl}\ }\textbf {\bibinfo {volume} {625}},\ \bibinfo {pages}
  {L95} (\bibinfo {year} {2005})},\ \Eprint
  {http://arxiv.org/abs/arXiv:astro-ph/0504450} {arXiv:astro-ph/0504450}
  \BibitemShut {NoStop}%
\bibitem [{\citenamefont {{Rees}}\ and\ \citenamefont
  {{M{\'e}sz{\'a}ros}}(2005)}]{Rees+05phot}%
  \BibitemOpen
  \bibfield  {author} {\bibinfo {author} {\bibfnamefont {M.~J.}\ \bibnamefont
  {{Rees}}}\ and\ \bibinfo {author} {\bibfnamefont {P.}~\bibnamefont
  {{M{\'e}sz{\'a}ros}}},\ }\href {\doibase 10.1086/430818} {\bibfield
  {journal} {\bibinfo  {journal} {Astrophys.J.}\ }\textbf {\bibinfo {volume}
  {628}},\ \bibinfo {pages} {847} (\bibinfo {year} {2005})},\ \Eprint
  {http://arxiv.org/abs/arXiv:astro-ph/0412702} {arXiv:astro-ph/0412702}
  \BibitemShut {NoStop}%
\bibitem [{\citenamefont {{Pe'er}}\ \emph {et~al.}(2006)\citenamefont
  {{Pe'er}}, \citenamefont {{M{\'e}sz{\'a}ros}},\ and\ \citenamefont
  {{Rees}}}]{Peer+06phot}%
  \BibitemOpen
  \bibfield  {author} {\bibinfo {author} {\bibfnamefont {A.}~\bibnamefont
  {{Pe'er}}}, \bibinfo {author} {\bibfnamefont {P.}~\bibnamefont
  {{M{\'e}sz{\'a}ros}}}, \ and\ \bibinfo {author} {\bibfnamefont {M.~J.}\
  \bibnamefont {{Rees}}},\ }\href {\doibase 10.1086/501424} {\bibfield
  {journal} {\bibinfo  {journal} {\apj}\ }\textbf {\bibinfo {volume} {642}},\
  \bibinfo {pages} {995} (\bibinfo {year} {2006})},\ \Eprint
  {http://arxiv.org/abs/arXiv:astro-ph/0510114} {arXiv:astro-ph/0510114}
  \BibitemShut {NoStop}%
\bibitem [{\citenamefont {{Beloborodov}}(2010)}]{Beloborodov10pn}%
  \BibitemOpen
  \bibfield  {author} {\bibinfo {author} {\bibfnamefont {A.~M.}\ \bibnamefont
  {{Beloborodov}}},\ }\href {\doibase 10.1111/j.1365-2966.2010.16770.x}
  {\bibfield  {journal} {\bibinfo  {journal} {\mnras}\ }\textbf {\bibinfo
  {volume} {407}},\ \bibinfo {pages} {1033} (\bibinfo {year} {2010})},\ \Eprint
  {http://arxiv.org/abs/0907.0732} {arXiv:0907.0732 [astro-ph.HE]} \BibitemShut
  {NoStop}%
\bibitem [{\citenamefont {{Pe'er}}(2011)}]{Peer11-fermigrb}%
  \BibitemOpen
  \bibfield  {author} {\bibinfo {author} {\bibfnamefont {A.}~\bibnamefont
  {{Pe'er}}},\ }\href@noop {} {\bibfield  {journal} {\bibinfo  {journal} {ArXiv
  e-prints}\ } (\bibinfo {year} {2011})},\ \Eprint
  {http://arxiv.org/abs/1111.3378} {arXiv:1111.3378 [astro-ph.HE]} \BibitemShut
  {NoStop}%
\bibitem [{\citenamefont {{Drenkhahn}}(2002)}]{Drenkhahn02}%
  \BibitemOpen
  \bibfield  {author} {\bibinfo {author} {\bibfnamefont {G.}~\bibnamefont
  {{Drenkhahn}}},\ }\href {\doibase 10.1051/0004-6361:20020390} {\bibfield
  {journal} {\bibinfo  {journal} {\aap}\ }\textbf {\bibinfo {volume} {387}},\
  \bibinfo {pages} {714} (\bibinfo {year} {2002})},\ \Eprint
  {http://arxiv.org/abs/arXiv:astro-ph/0112509} {arXiv:astro-ph/0112509}
  \BibitemShut {NoStop}%
\bibitem [{\citenamefont {{Tchekhovskoy}}\ \emph {et~al.}(2010)\citenamefont
  {{Tchekhovskoy}}, \citenamefont {{Narayan}},\ and\ \citenamefont
  {{McKinney}}}]{Tchekhovskoy+10grb}%
  \BibitemOpen
  \bibfield  {author} {\bibinfo {author} {\bibfnamefont {A.}~\bibnamefont
  {{Tchekhovskoy}}}, \bibinfo {author} {\bibfnamefont {R.}~\bibnamefont
  {{Narayan}}}, \ and\ \bibinfo {author} {\bibfnamefont {J.~C.}\ \bibnamefont
  {{McKinney}}},\ }\href {\doibase 10.1016/j.newast.2010.03.001} {\bibfield
  {journal} {\bibinfo  {journal} {\na}\ }\textbf {\bibinfo {volume} {15}},\
  \bibinfo {pages} {749} (\bibinfo {year} {2010})},\ \Eprint
  {http://arxiv.org/abs/0909.0011} {arXiv:0909.0011 [astro-ph.HE]} \BibitemShut
  {NoStop}%
\bibitem [{\citenamefont {{Metzger}}\ \emph {et~al.}(2011)\citenamefont
  {{Metzger}}, \citenamefont {{Giannios}}, \citenamefont {{Thompson}},
  \citenamefont {{Bucciantini}},\ and\ \citenamefont
  {{Quataert}}}]{Metzger+11grbmag}%
  \BibitemOpen
  \bibfield  {author} {\bibinfo {author} {\bibfnamefont {B.~D.}\ \bibnamefont
  {{Metzger}}}, \bibinfo {author} {\bibfnamefont {D.}~\bibnamefont
  {{Giannios}}}, \bibinfo {author} {\bibfnamefont {T.~A.}\ \bibnamefont
  {{Thompson}}}, \bibinfo {author} {\bibfnamefont {N.}~\bibnamefont
  {{Bucciantini}}}, \ and\ \bibinfo {author} {\bibfnamefont {E.}~\bibnamefont
  {{Quataert}}},\ }\href {\doibase 10.1111/j.1365-2966.2011.18280.x} {\bibfield
   {journal} {\bibinfo  {journal} {\mnras}\ }\textbf {\bibinfo {volume}
  {413}},\ \bibinfo {pages} {2031} (\bibinfo {year} {2011})},\ \Eprint
  {http://arxiv.org/abs/1012.0001} {arXiv:1012.0001 [astro-ph.HE]} \BibitemShut
  {NoStop}%
\bibitem [{\citenamefont {{McKinney}}\ and\ \citenamefont
  {{Uzdensky}}(2012)}]{McKinney+12magphot}%
  \BibitemOpen
  \bibfield  {author} {\bibinfo {author} {\bibfnamefont {J.~C.}\ \bibnamefont
  {{McKinney}}}\ and\ \bibinfo {author} {\bibfnamefont {D.~A.}\ \bibnamefont
  {{Uzdensky}}},\ }\href {\doibase 10.1111/j.1365-2966.2011.19721.x} {\bibfield
   {journal} {\bibinfo  {journal} {\mnras}\ }\textbf {\bibinfo {volume}
  {419}},\ \bibinfo {pages} {573} (\bibinfo {year} {2012})},\ \Eprint
  {http://arxiv.org/abs/1011.1904} {arXiv:1011.1904 [astro-ph.HE]} \BibitemShut
  {NoStop}%
\bibitem [{\citenamefont {{Giannios}}\ and\ \citenamefont
  {{Spruit}}(2007)}]{Giannios+07photspec}%
  \BibitemOpen
  \bibfield  {author} {\bibinfo {author} {\bibfnamefont {D.}~\bibnamefont
  {{Giannios}}}\ and\ \bibinfo {author} {\bibfnamefont {H.~C.}\ \bibnamefont
  {{Spruit}}},\ }\href {\doibase 10.1051/0004-6361:20066739} {\bibfield
  {journal} {\bibinfo  {journal} {\aap}\ }\textbf {\bibinfo {volume} {469}},\
  \bibinfo {pages} {1} (\bibinfo {year} {2007})},\ \Eprint
  {http://arxiv.org/abs/arXiv:astro-ph/0611385} {arXiv:astro-ph/0611385}
  \BibitemShut {NoStop}%
\bibitem [{\citenamefont {{Veres}}\ and\ \citenamefont
  {{M{\'e}sz{\'a}ros}}(2012)}]{Veres+12mag}%
  \BibitemOpen
  \bibfield  {author} {\bibinfo {author} {\bibfnamefont {P.}~\bibnamefont
  {{Veres}}}\ and\ \bibinfo {author} {\bibfnamefont {P.}~\bibnamefont
  {{M{\'e}sz{\'a}ros}}},\ }\href {\doibase 10.1088/0004-637X/755/1/12}
  {\bibfield  {journal} {\bibinfo  {journal} {\apj}\ }\textbf {\bibinfo
  {volume} {755}},\ \bibinfo {eid} {12} (\bibinfo {year} {2012})},\ \Eprint
  {http://arxiv.org/abs/1202.2821} {arXiv:1202.2821 [astro-ph.HE]} \BibitemShut
  {NoStop}%
\bibitem [{\citenamefont {{M{\'e}sz{\'a}ros}}\ and\ \citenamefont
  {{Gehrels}}(2012)}]{Meszaros+12raa}%
  \BibitemOpen
  \bibfield  {author} {\bibinfo {author} {\bibfnamefont {P.}~\bibnamefont
  {{M{\'e}sz{\'a}ros}}}\ and\ \bibinfo {author} {\bibfnamefont
  {N.}~\bibnamefont {{Gehrels}}},\ }\href {\doibase 10.1088/1674-4527/12/8/012}
  {\bibfield  {journal} {\bibinfo  {journal} {Research in Astronomy and
  Astrophysics}\ }\textbf {\bibinfo {volume} {12}},\ \bibinfo {pages} {1139}
  (\bibinfo {year} {2012})}\BibitemShut {NoStop}%
\bibitem [{\citenamefont {{M{\'e}sz{\'a}ros}}\ and\ \citenamefont
  {{Rees}}(2011{\natexlab{a}})}]{Meszaros+11gevmag}%
  \BibitemOpen
  \bibfield  {author} {\bibinfo {author} {\bibfnamefont {P.}~\bibnamefont
  {{M{\'e}sz{\'a}ros}}}\ and\ \bibinfo {author} {\bibfnamefont {M.~J.}\
  \bibnamefont {{Rees}}},\ }\href {\doibase 10.1088/2041-8205/733/2/L40}
  {\bibfield  {journal} {\bibinfo  {journal} {\apjl}\ }\textbf {\bibinfo
  {volume} {733}},\ \bibinfo {pages} {L40+} (\bibinfo {year}
  {2011}{\natexlab{a}})},\ \Eprint {http://arxiv.org/abs/1104.5025}
  {arXiv:1104.5025 [astro-ph.HE]} \BibitemShut {NoStop}%
\bibitem [{\citenamefont {{Thompson}}(1994)}]{Thompson94}%
  \BibitemOpen
  \bibfield  {author} {\bibinfo {author} {\bibfnamefont {C.}~\bibnamefont
  {{Thompson}}},\ }\href@noop {} {\bibfield  {journal} {\bibinfo  {journal}
  {\mnras}\ }\textbf {\bibinfo {volume} {270}},\ \bibinfo {pages} {480}
  (\bibinfo {year} {1994})}\BibitemShut {NoStop}%
\bibitem [{\citenamefont {{Bo{\v s}njak}}\ and\ \citenamefont
  {{Kumar}}(2012)}]{Bosnjak+12delay}%
  \BibitemOpen
  \bibfield  {author} {\bibinfo {author} {\bibfnamefont {{\v Z}.}~\bibnamefont
  {{Bo{\v s}njak}}}\ and\ \bibinfo {author} {\bibfnamefont {P.}~\bibnamefont
  {{Kumar}}},\ }\href {\doibase 10.1111/j.1745-3933.2011.01202.x} {\bibfield
  {journal} {\bibinfo  {journal} {\mnras}\ }\textbf {\bibinfo {volume} {421}},\
  \bibinfo {pages} {L39} (\bibinfo {year} {2012})},\ \Eprint
  {http://arxiv.org/abs/1108.0929} {arXiv:1108.0929 [astro-ph.HE]} \BibitemShut
  {NoStop}%
\bibitem [{\citenamefont {{Bahcall}}\ and\ \citenamefont
  {{M{\'e}sz{\'a}ros}}(2000)}]{Bahcall+00pn}%
  \BibitemOpen
  \bibfield  {author} {\bibinfo {author} {\bibfnamefont {J.~N.}\ \bibnamefont
  {{Bahcall}}}\ and\ \bibinfo {author} {\bibfnamefont {P.}~\bibnamefont
  {{M{\'e}sz{\'a}ros}}},\ }\href@noop {} {\bibfield  {journal} {\bibinfo
  {journal} {Physical Review Letters}\ }\textbf {\bibinfo {volume} {85}},\
  \bibinfo {pages} {1362} (\bibinfo {year} {2000})},\ \Eprint
  {http://arxiv.org/abs/arXiv:hep-ph/0004019} {arXiv:hep-ph/0004019}
  \BibitemShut {NoStop}%
\bibitem [{\citenamefont {{Band}}\ \emph {et~al.}(1993)\citenamefont {{Band}},
  \citenamefont {{Matteson}}, \citenamefont {{Ford}}, \citenamefont
  {{Schaefer}}, \citenamefont {{Palmer}}, \citenamefont {{Teegarden}},
  \citenamefont {{Cline}}, \citenamefont {{Briggs}}, \citenamefont
  {{Paciesas}}, \citenamefont {{Pendleton}}, \citenamefont {{Fishman}},
  \citenamefont {{Kouveliotou}}, \citenamefont {{Meegan}}, \citenamefont
  {{Wilson}},\ and\ \citenamefont {{Lestrade}}}]{Band+93}%
  \BibitemOpen
  \bibfield  {author} {\bibinfo {author} {\bibfnamefont {D.}~\bibnamefont
  {{Band}}}, \bibinfo {author} {\bibfnamefont {J.}~\bibnamefont {{Matteson}}},
  \bibinfo {author} {\bibfnamefont {L.}~\bibnamefont {{Ford}}}, \bibinfo
  {author} {\bibfnamefont {B.}~\bibnamefont {{Schaefer}}}, \bibinfo {author}
  {\bibfnamefont {D.}~\bibnamefont {{Palmer}}}, \bibinfo {author}
  {\bibfnamefont {B.}~\bibnamefont {{Teegarden}}}, \bibinfo {author}
  {\bibfnamefont {T.}~\bibnamefont {{Cline}}}, \bibinfo {author} {\bibfnamefont
  {M.}~\bibnamefont {{Briggs}}}, \bibinfo {author} {\bibfnamefont
  {W.}~\bibnamefont {{Paciesas}}}, \bibinfo {author} {\bibfnamefont
  {G.}~\bibnamefont {{Pendleton}}}, \bibinfo {author} {\bibfnamefont
  {G.}~\bibnamefont {{Fishman}}}, \bibinfo {author} {\bibfnamefont
  {C.}~\bibnamefont {{Kouveliotou}}}, \bibinfo {author} {\bibfnamefont
  {C.}~\bibnamefont {{Meegan}}}, \bibinfo {author} {\bibfnamefont
  {R.}~\bibnamefont {{Wilson}}}, \ and\ \bibinfo {author} {\bibfnamefont
  {P.}~\bibnamefont {{Lestrade}}},\ }\href {\doibase 10.1086/172995} {\bibfield
   {journal} {\bibinfo  {journal} {\apj}\ }\textbf {\bibinfo {volume} {413}},\
  \bibinfo {pages} {281} (\bibinfo {year} {1993})}\BibitemShut {NoStop}%
\bibitem [{\citenamefont {{Toma}}\ \emph {et~al.}(2011)\citenamefont {{Toma}},
  \citenamefont {{Sakamoto}},\ and\ \citenamefont
  {{M{\'e}sz{\'a}ros}}}]{Toma+11pop3}%
  \BibitemOpen
  \bibfield  {author} {\bibinfo {author} {\bibfnamefont {K.}~\bibnamefont
  {{Toma}}}, \bibinfo {author} {\bibfnamefont {T.}~\bibnamefont {{Sakamoto}}},
  \ and\ \bibinfo {author} {\bibfnamefont {P.}~\bibnamefont
  {{M{\'e}sz{\'a}ros}}},\ }\href {\doibase 10.1088/0004-637X/731/2/127}
  {\bibfield  {journal} {\bibinfo  {journal} {\apj}\ }\textbf {\bibinfo
  {volume} {731}},\ \bibinfo {pages} {127} (\bibinfo {year} {2011})},\ \Eprint
  {http://arxiv.org/abs/1008.1269} {arXiv:1008.1269 [astro-ph.CO]} \BibitemShut
  {NoStop}%
\bibitem [{\citenamefont {{Asano}}\ \emph {et~al.}(2011)\citenamefont
  {{Asano}}, \citenamefont {{M{\'e}sz{\'a}ros}}, \citenamefont {{Murase}},
  \citenamefont {{Inoue}},\ and\ \citenamefont
  {{Terasawa}}}]{Asano+11fermiextra}%
  \BibitemOpen
  \bibfield  {author} {\bibinfo {author} {\bibfnamefont {K.}~\bibnamefont
  {{Asano}}}, \bibinfo {author} {\bibfnamefont {P.}~\bibnamefont
  {{M{\'e}sz{\'a}ros}}}, \bibinfo {author} {\bibfnamefont {K.}~\bibnamefont
  {{Murase}}}, \bibinfo {author} {\bibfnamefont {S.}~\bibnamefont {{Inoue}}}, \
  and\ \bibinfo {author} {\bibfnamefont {T.}~\bibnamefont {{Terasawa}}},\
  }\href@noop {} {\bibfield  {journal} {\bibinfo  {journal} {ArXiv e-prints}\ }
  (\bibinfo {year} {2011})},\ \Eprint {http://arxiv.org/abs/1111.0127}
  {arXiv:1111.0127 [astro-ph.HE]} \BibitemShut {NoStop}%
\bibitem [{\citenamefont {{Kowal}}\ \emph {et~al.}(2009)\citenamefont
  {{Kowal}}, \citenamefont {{Lazarian}}, \citenamefont {{Vishniac}},\ and\
  \citenamefont {{Otmianowska-Mazur}}}]{Kowal+09FastRC}%
  \BibitemOpen
  \bibfield  {author} {\bibinfo {author} {\bibfnamefont {G.}~\bibnamefont
  {{Kowal}}}, \bibinfo {author} {\bibfnamefont {A.}~\bibnamefont {{Lazarian}}},
  \bibinfo {author} {\bibfnamefont {E.~T.}\ \bibnamefont {{Vishniac}}}, \ and\
  \bibinfo {author} {\bibfnamefont {K.}~\bibnamefont {{Otmianowska-Mazur}}},\
  }\href {\doibase 10.1088/0004-637X/700/1/63} {\bibfield  {journal} {\bibinfo
  {journal} {\apj}\ }\textbf {\bibinfo {volume} {700}},\ \bibinfo {pages} {63}
  (\bibinfo {year} {2009})},\ \Eprint {http://arxiv.org/abs/0903.2052}
  {arXiv:0903.2052 [astro-ph.GA]} \BibitemShut {NoStop}%
\bibitem [{\citenamefont {{Sironi}}\ and\ \citenamefont
  {{Spitkovsky}}(2011)}]{Sironi+11}%
  \BibitemOpen
  \bibfield  {author} {\bibinfo {author} {\bibfnamefont {L.}~\bibnamefont
  {{Sironi}}}\ and\ \bibinfo {author} {\bibfnamefont {A.}~\bibnamefont
  {{Spitkovsky}}},\ }\href {\doibase 10.1088/0004-637X/741/1/39} {\bibfield
  {journal} {\bibinfo  {journal} {\apj}\ }\textbf {\bibinfo {volume} {741}},\
  \bibinfo {eid} {39} (\bibinfo {year} {2011})},\ \Eprint
  {http://arxiv.org/abs/1107.0977} {arXiv:1107.0977 [astro-ph.HE]} \BibitemShut
  {NoStop}%
\bibitem [{\citenamefont {{Hoshino}}(2012)}]{Hoshino+12}%
  \BibitemOpen
  \bibfield  {author} {\bibinfo {author} {\bibfnamefont {M.}~\bibnamefont
  {{Hoshino}}},\ }\href {\doibase 10.1103/PhysRevLett.108.135003} {\bibfield
  {journal} {\bibinfo  {journal} {Physical Review Letters}\ }\textbf {\bibinfo
  {volume} {108}},\ \bibinfo {eid} {135003} (\bibinfo {year} {2012})},\ \Eprint
  {http://arxiv.org/abs/1201.0837} {arXiv:1201.0837 [astro-ph.HE]} \BibitemShut
  {NoStop}%
\bibitem [{\citenamefont {{Dermer}}\ and\ \citenamefont
  {{Menon}}(2009)}]{Dermer+09book}%
  \BibitemOpen
  \bibfield  {author} {\bibinfo {author} {\bibfnamefont {C.~D.}\ \bibnamefont
  {{Dermer}}}\ and\ \bibinfo {author} {\bibfnamefont {G.}~\bibnamefont
  {{Menon}}},\ }\href@noop {} {\emph {\bibinfo {title} {High Energy Radiation
  from Black Holes: Gamma Rays, Cosmic Rays, and Neutrinos by Charles D.~Dermer
  and Govind Menon.~Princeton Univerisity Press, November 2009.}}},\ edited by\
  \bibinfo {editor} {\bibnamefont {{Dermer, C.~D.~\& Menon, G.}}}\ (\bibinfo
  {year} {2009})\BibitemShut {NoStop}%
\bibitem [{\citenamefont {{Kamae}}\ \emph {et~al.}(2006)\citenamefont
  {{Kamae}}, \citenamefont {{Karlsson}}, \citenamefont {{Mizuno}},
  \citenamefont {{Abe}},\ and\ \citenamefont {{Koi}}}]{Kamae+06}%
  \BibitemOpen
  \bibfield  {author} {\bibinfo {author} {\bibfnamefont {T.}~\bibnamefont
  {{Kamae}}}, \bibinfo {author} {\bibfnamefont {N.}~\bibnamefont {{Karlsson}}},
  \bibinfo {author} {\bibfnamefont {T.}~\bibnamefont {{Mizuno}}}, \bibinfo
  {author} {\bibfnamefont {T.}~\bibnamefont {{Abe}}}, \ and\ \bibinfo {author}
  {\bibfnamefont {T.}~\bibnamefont {{Koi}}},\ }\href {\doibase 10.1086/505189}
  {\bibfield  {journal} {\bibinfo  {journal} {\apj}\ }\textbf {\bibinfo
  {volume} {647}},\ \bibinfo {pages} {692} (\bibinfo {year} {2006})},\ \Eprint
  {http://arxiv.org/abs/arXiv:astro-ph/0605581} {arXiv:astro-ph/0605581}
  \BibitemShut {NoStop}%
\bibitem [{\citenamefont {{Gao}}\ and\ \citenamefont
  {{M{\'e}sz{\'a}ros}}(2012)}]{Gao+12magnu}%
  \BibitemOpen
  \bibfield  {author} {\bibinfo {author} {\bibfnamefont {S.}~\bibnamefont
  {{Gao}}}\ and\ \bibinfo {author} {\bibfnamefont {P.}~\bibnamefont
  {{M{\'e}sz{\'a}ros}}},\ }\href {\doibase 10.1103/PhysRevD.85.103009}
  {\bibfield  {journal} {\bibinfo  {journal} {\prd}\ }\textbf {\bibinfo
  {volume} {85}},\ \bibinfo {eid} {103009} (\bibinfo {year} {2012})},\ \Eprint
  {http://arxiv.org/abs/1112.5664} {arXiv:1112.5664 [astro-ph.HE]} \BibitemShut
  {NoStop}%
\bibitem [{\citenamefont {{Jones}}(1965)}]{Jones+65}%
  \BibitemOpen
  \bibfield  {author} {\bibinfo {author} {\bibfnamefont {F.~C.}\ \bibnamefont
  {{Jones}}},\ }\href {\doibase 10.1103/PhysRev.137.B1306} {\bibfield
  {journal} {\bibinfo  {journal} {Physical Review}\ }\textbf {\bibinfo {volume}
  {137}},\ \bibinfo {pages} {1306} (\bibinfo {year} {1965})}\BibitemShut
  {NoStop}%
\bibitem [{\citenamefont {{Asano}}\ and\ \citenamefont
  {{Nagataki}}(2006)}]{Asano+06}%
  \BibitemOpen
  \bibfield  {author} {\bibinfo {author} {\bibfnamefont {K.}~\bibnamefont
  {{Asano}}}\ and\ \bibinfo {author} {\bibfnamefont {S.}~\bibnamefont
  {{Nagataki}}},\ }\href {\doibase 10.1086/503291} {\bibfield  {journal}
  {\bibinfo  {journal} {\apjl}\ }\textbf {\bibinfo {volume} {640}},\ \bibinfo
  {pages} {L9} (\bibinfo {year} {2006})},\ \Eprint
  {http://arxiv.org/abs/arXiv:astro-ph/0603107} {arXiv:astro-ph/0603107}
  \BibitemShut {NoStop}%
\bibitem [{\citenamefont {{Ando}}\ and\ \citenamefont
  {{Beacom}}(2005)}]{Ando+05}%
  \BibitemOpen
  \bibfield  {author} {\bibinfo {author} {\bibfnamefont {S.}~\bibnamefont
  {{Ando}}}\ and\ \bibinfo {author} {\bibfnamefont {J.~F.}\ \bibnamefont
  {{Beacom}}},\ }\href {\doibase 10.1103/PhysRevLett.95.061103} {\bibfield
  {journal} {\bibinfo  {journal} {Physical Review Letters}\ }\textbf {\bibinfo
  {volume} {95}},\ \bibinfo {eid} {061103} (\bibinfo {year} {2005})},\ \Eprint
  {http://arxiv.org/abs/arXiv:astro-ph/0502521} {arXiv:astro-ph/0502521}
  \BibitemShut {NoStop}%
\bibitem [{\citenamefont {{Marscher}}\ \emph {et~al.}(1980)\citenamefont
  {{Marscher}}, \citenamefont {{Vestrand}},\ and\ \citenamefont
  {{Scott}}}]{Marscher+80}%
  \BibitemOpen
  \bibfield  {author} {\bibinfo {author} {\bibfnamefont {A.~P.}\ \bibnamefont
  {{Marscher}}}, \bibinfo {author} {\bibfnamefont {W.~T.}\ \bibnamefont
  {{Vestrand}}}, \ and\ \bibinfo {author} {\bibfnamefont {J.~S.}\ \bibnamefont
  {{Scott}}},\ }\href {\doibase 10.1086/158433} {\bibfield  {journal} {\bibinfo
   {journal} {\apj}\ }\textbf {\bibinfo {volume} {241}},\ \bibinfo {pages}
  {1166} (\bibinfo {year} {1980})}\BibitemShut {NoStop}%
\bibitem [{\citenamefont {{H{\"u}mmer}}\ \emph {et~al.}(2012)\citenamefont
  {{H{\"u}mmer}}, \citenamefont {{Baerwald}},\ and\ \citenamefont
  {{Winter}}}]{Hummer+11nu-ic3}%
  \BibitemOpen
  \bibfield  {author} {\bibinfo {author} {\bibfnamefont {S.}~\bibnamefont
  {{H{\"u}mmer}}}, \bibinfo {author} {\bibfnamefont {P.}~\bibnamefont
  {{Baerwald}}}, \ and\ \bibinfo {author} {\bibfnamefont {W.}~\bibnamefont
  {{Winter}}},\ }\href {\doibase 10.1103/PhysRevLett.108.231101} {\bibfield
  {journal} {\bibinfo  {journal} {Physical Review Letters}\ }\textbf {\bibinfo
  {volume} {108}},\ \bibinfo {eid} {231101} (\bibinfo {year} {2012})},\ \Eprint
  {http://arxiv.org/abs/1112.1076} {arXiv:1112.1076 [astro-ph.HE]} \BibitemShut
  {NoStop}%
\bibitem [{\citenamefont {{Asano}}\ and\ \citenamefont
  {{M{\'e}sz{\'a}ros}}(2012)}]{Asano+12grbhad}%
  \BibitemOpen
  \bibfield  {author} {\bibinfo {author} {\bibfnamefont {K.}~\bibnamefont
  {{Asano}}}\ and\ \bibinfo {author} {\bibfnamefont {P.}~\bibnamefont
  {{M{\'e}sz{\'a}ros}}},\ }\href@noop {} {\bibfield  {journal} {\bibinfo
  {journal} {ArXiv e-prints}\ } (\bibinfo {year} {2012})},\ \Eprint
  {http://arxiv.org/abs/1206.0347} {arXiv:1206.0347 [astro-ph.HE]} \BibitemShut
  {NoStop}%
\bibitem [{\citenamefont {{Drury}}(2012)}]{Drury+12}%
  \BibitemOpen
  \bibfield  {author} {\bibinfo {author} {\bibfnamefont {L.~O.}\ \bibnamefont
  {{Drury}}},\ }\href {\doibase 10.1111/j.1365-2966.2012.20804.x} {\bibfield
  {journal} {\bibinfo  {journal} {\mnras}\ }\textbf {\bibinfo {volume} {422}},\
  \bibinfo {pages} {2474} (\bibinfo {year} {2012})},\ \Eprint
  {http://arxiv.org/abs/1201.6612} {arXiv:1201.6612 [astro-ph.HE]} \BibitemShut
  {NoStop}%
\bibitem [{\citenamefont {{Bosch-Ramon}}(2012)}]{Bosch-Ramon+12}%
  \BibitemOpen
  \bibfield  {author} {\bibinfo {author} {\bibfnamefont {V.}~\bibnamefont
  {{Bosch-Ramon}}},\ }\href {\doibase 10.1051/0004-6361/201219231} {\bibfield
  {journal} {\bibinfo  {journal} {\aap}\ }\textbf {\bibinfo {volume} {542}},\
  \bibinfo {eid} {A125} (\bibinfo {year} {2012})},\ \Eprint
  {http://arxiv.org/abs/1205.3450} {arXiv:1205.3450 [astro-ph.HE]} \BibitemShut
  {NoStop}%
\bibitem [{\citenamefont {{Wanderman}}\ and\ \citenamefont
  {{Piran}}(2010)}]{Wanderman+10grbsfr}%
  \BibitemOpen
  \bibfield  {author} {\bibinfo {author} {\bibfnamefont {D.}~\bibnamefont
  {{Wanderman}}}\ and\ \bibinfo {author} {\bibfnamefont {T.}~\bibnamefont
  {{Piran}}},\ }\href {\doibase 10.1111/j.1365-2966.2010.16787.x} {\bibfield
  {journal} {\bibinfo  {journal} {\mnras}\ }\textbf {\bibinfo {volume} {406}},\
  \bibinfo {pages} {1944} (\bibinfo {year} {2010})},\ \Eprint
  {http://arxiv.org/abs/0912.0709} {arXiv:0912.0709 [astro-ph.HE]} \BibitemShut
  {NoStop}%
\bibitem [{\citenamefont {{Liang}}\ \emph {et~al.}(2007)\citenamefont
  {{Liang}}, \citenamefont {{Zhang}}, \citenamefont {{Virgili}},\ and\
  \citenamefont {{Dai}}}]{Liang+07lowlumgrb}%
  \BibitemOpen
  \bibfield  {author} {\bibinfo {author} {\bibfnamefont {E.}~\bibnamefont
  {{Liang}}}, \bibinfo {author} {\bibfnamefont {B.}~\bibnamefont {{Zhang}}},
  \bibinfo {author} {\bibfnamefont {F.}~\bibnamefont {{Virgili}}}, \ and\
  \bibinfo {author} {\bibfnamefont {Z.~G.}\ \bibnamefont {{Dai}}},\ }\href
  {\doibase 10.1086/517959} {\bibfield  {journal} {\bibinfo  {journal} {\apj}\
  }\textbf {\bibinfo {volume} {662}},\ \bibinfo {pages} {1111} (\bibinfo {year}
  {2007})},\ \Eprint {http://arxiv.org/abs/arXiv:astro-ph/0605200}
  {arXiv:astro-ph/0605200} \BibitemShut {NoStop}%
\bibitem [{\citenamefont {{Blandford}}\ and\ \citenamefont
  {{Znajek}}(1977)}]{Blandford+77znajek}%
  \BibitemOpen
  \bibfield  {author} {\bibinfo {author} {\bibfnamefont {R.~D.}\ \bibnamefont
  {{Blandford}}}\ and\ \bibinfo {author} {\bibfnamefont {R.~L.}\ \bibnamefont
  {{Znajek}}},\ }\href@noop {} {\bibfield  {journal} {\bibinfo  {journal}
  {\mnras}\ }\textbf {\bibinfo {volume} {179}},\ \bibinfo {pages} {433}
  (\bibinfo {year} {1977})}\BibitemShut {NoStop}%
\bibitem [{\citenamefont {{Drenkhahn}}\ and\ \citenamefont
  {{Spruit}}(2002)}]{Drenkhahn+02}%
  \BibitemOpen
  \bibfield  {author} {\bibinfo {author} {\bibfnamefont {G.}~\bibnamefont
  {{Drenkhahn}}}\ and\ \bibinfo {author} {\bibfnamefont {H.~C.}\ \bibnamefont
  {{Spruit}}},\ }\href {\doibase 10.1051/0004-6361:20020839} {\bibfield
  {journal} {\bibinfo  {journal} {\aap}\ }\textbf {\bibinfo {volume} {391}},\
  \bibinfo {pages} {1141} (\bibinfo {year} {2002})},\ \Eprint
  {http://arxiv.org/abs/arXiv:astro-ph/0202387} {arXiv:astro-ph/0202387}
  \BibitemShut {NoStop}%
\bibitem [{\citenamefont {{Asano}}\ and\ \citenamefont
  {{M{\'e}sz{\'a}ros}}(2011)}]{Asano+11grbtemp}%
  \BibitemOpen
  \bibfield  {author} {\bibinfo {author} {\bibfnamefont {K.}~\bibnamefont
  {{Asano}}}\ and\ \bibinfo {author} {\bibfnamefont {P.}~\bibnamefont
  {{M{\'e}sz{\'a}ros}}},\ }\href {\doibase 10.1088/0004-637X/739/2/103}
  {\bibfield  {journal} {\bibinfo  {journal} {\apj}\ }\textbf {\bibinfo
  {volume} {739}},\ \bibinfo {eid} {103} (\bibinfo {year} {2011})},\ \Eprint
  {http://arxiv.org/abs/1107.4825} {arXiv:1107.4825 [astro-ph.HE]} \BibitemShut
  {NoStop}%
\bibitem [{\citenamefont {{M{\'e}sz{\'a}ros}}\ and\ \citenamefont
  {{Rees}}(2011{\natexlab{b}})}]{Meszaros+11col}%
  \BibitemOpen
  \bibfield  {author} {\bibinfo {author} {\bibfnamefont {P.}~\bibnamefont
  {{M{\'e}sz{\'a}ros}}}\ and\ \bibinfo {author} {\bibfnamefont {M.~J.}\
  \bibnamefont {{Rees}}},\ }\href@noop {} {\bibfield  {journal} {\bibinfo
  {journal} {\apjl}\ }\textbf {\bibinfo {volume} {733}},\ \bibinfo {pages}
  {L40+} (\bibinfo {year} {2011}{\natexlab{b}})},\ \Eprint
  {http://arxiv.org/abs/1104.5025} {arXiv:1104.5025 [astro-ph.HE]} \BibitemShut
  {NoStop}%
\bibitem [{\citenamefont {{Murase}}(2008)}]{Murase+08photonu}%
  \BibitemOpen
  \bibfield  {author} {\bibinfo {author} {\bibfnamefont {K.}~\bibnamefont
  {{Murase}}},\ }\href {\doibase 10.1103/PhysRevD.78.101302} {\bibfield
  {journal} {\bibinfo  {journal} {\prd}\ }\textbf {\bibinfo {volume} {78}},\
  \bibinfo {eid} {101302} (\bibinfo {year} {2008})},\ \Eprint
  {http://arxiv.org/abs/0807.0919} {arXiv:0807.0919} \BibitemShut {NoStop}%
\bibitem [{\citenamefont {{Wang}}\ and\ \citenamefont
  {{Dai}}(2009)}]{Wang+09photonu}%
  \BibitemOpen
  \bibfield  {author} {\bibinfo {author} {\bibfnamefont {X.-Y.}\ \bibnamefont
  {{Wang}}}\ and\ \bibinfo {author} {\bibfnamefont {Z.-G.}\ \bibnamefont
  {{Dai}}},\ }\href {\doibase 10.1088/0004-637X/691/2/L67} {\bibfield
  {journal} {\bibinfo  {journal} {\apjl}\ }\textbf {\bibinfo {volume} {691}},\
  \bibinfo {pages} {L67} (\bibinfo {year} {2009})},\ \Eprint
  {http://arxiv.org/abs/0807.0290} {arXiv:0807.0290} \BibitemShut {NoStop}%
\bibitem [{\citenamefont {{Zhang}}\ and\ \citenamefont
  {{Kumar}}(2012)}]{Zhang+12grbnu}%
  \BibitemOpen
  \bibfield  {author} {\bibinfo {author} {\bibfnamefont {B.}~\bibnamefont
  {{Zhang}}}\ and\ \bibinfo {author} {\bibfnamefont {P.}~\bibnamefont
  {{Kumar}}},\ }\href@noop {} {\bibfield  {journal} {\bibinfo  {journal} {ArXiv
  e-prints}\ } (\bibinfo {year} {2012})},\ \Eprint
  {http://arxiv.org/abs/1210.0647} {arXiv:1210.0647 [astro-ph.HE]} \BibitemShut
  {NoStop}%
\bibitem [{\citenamefont {{Guetta}}\ \emph
  {et~al.}(2001{\natexlab{a}})\citenamefont {{Guetta}}, \citenamefont
  {{Spada}},\ and\ \citenamefont {{Waxman}}}]{Guetta+01grbnu}%
  \BibitemOpen
  \bibfield  {author} {\bibinfo {author} {\bibfnamefont {D.}~\bibnamefont
  {{Guetta}}}, \bibinfo {author} {\bibfnamefont {M.}~\bibnamefont {{Spada}}}, \
  and\ \bibinfo {author} {\bibfnamefont {E.}~\bibnamefont {{Waxman}}},\ }\href
  {\doibase 10.1086/322481} {\bibfield  {journal} {\bibinfo  {journal} {\apj}\
  }\textbf {\bibinfo {volume} {559}},\ \bibinfo {pages} {101} (\bibinfo {year}
  {2001}{\natexlab{a}})},\ \Eprint
  {http://arxiv.org/abs/arXiv:astro-ph/0102487} {arXiv:astro-ph/0102487}
  \BibitemShut {NoStop}%
\bibitem [{\citenamefont {{Guetta}}\ \emph
  {et~al.}(2001{\natexlab{b}})\citenamefont {{Guetta}}, \citenamefont
  {{Spada}},\ and\ \citenamefont {{Waxman}}}]{Guetta+01is}%
  \BibitemOpen
  \bibfield  {author} {\bibinfo {author} {\bibfnamefont {D.}~\bibnamefont
  {{Guetta}}}, \bibinfo {author} {\bibfnamefont {M.}~\bibnamefont {{Spada}}}, \
  and\ \bibinfo {author} {\bibfnamefont {E.}~\bibnamefont {{Waxman}}},\ }\href
  {\doibase 10.1086/321543} {\bibfield  {journal} {\bibinfo  {journal} {\apj}\
  }\textbf {\bibinfo {volume} {557}},\ \bibinfo {pages} {399} (\bibinfo {year}
  {2001}{\natexlab{b}})},\ \Eprint
  {http://arxiv.org/abs/arXiv:astro-ph/0011170} {arXiv:astro-ph/0011170}
  \BibitemShut {NoStop}%
\bibitem [{\citenamefont {{Kotera}}\ and\ \citenamefont
  {{Olinto}}(2011)}]{Kotera+11uhecr}%
  \BibitemOpen
  \bibfield  {author} {\bibinfo {author} {\bibfnamefont {K.}~\bibnamefont
  {{Kotera}}}\ and\ \bibinfo {author} {\bibfnamefont {A.~V.}\ \bibnamefont
  {{Olinto}}},\ }\href {\doibase 10.1146/annurev-astro-081710-102620}
  {\bibfield  {journal} {\bibinfo  {journal} {\araa}\ }\textbf {\bibinfo
  {volume} {49}},\ \bibinfo {pages} {119} (\bibinfo {year} {2011})},\ \Eprint
  {http://arxiv.org/abs/1101.4256} {arXiv:1101.4256 [astro-ph.HE]} \BibitemShut
  {NoStop}%
\bibitem [{\citenamefont {{Ahlers}}\ \emph {et~al.}(2011)\citenamefont
  {{Ahlers}}, \citenamefont {{Gonzalez-Garcia}},\ and\ \citenamefont
  {{Halzen}}}]{Ahlers+11-grbprob}%
  \BibitemOpen
  \bibfield  {author} {\bibinfo {author} {\bibfnamefont {M.}~\bibnamefont
  {{Ahlers}}}, \bibinfo {author} {\bibfnamefont {M.~C.}\ \bibnamefont
  {{Gonzalez-Garcia}}}, \ and\ \bibinfo {author} {\bibfnamefont
  {F.}~\bibnamefont {{Halzen}}},\ }\href {\doibase
  10.1016/j.astropartphys.2011.05.008} {\bibfield  {journal} {\bibinfo
  {journal} {Astroparticle Physics}\ }\textbf {\bibinfo {volume} {35}},\
  \bibinfo {pages} {87} (\bibinfo {year} {2011})},\ \Eprint
  {http://arxiv.org/abs/1103.3421} {arXiv:1103.3421 [astro-ph.HE]} \BibitemShut
  {NoStop}%
\bibitem [{\citenamefont {{Li}}(2012)}]{Li11-nu-ic3}%
  \BibitemOpen
  \bibfield  {author} {\bibinfo {author} {\bibfnamefont {Z.}~\bibnamefont
  {{Li}}},\ }\href {\doibase 10.1103/PhysRevD.85.027301} {\bibfield  {journal}
  {\bibinfo  {journal} {\prd}\ }\textbf {\bibinfo {volume} {85}},\ \bibinfo
  {eid} {027301} (\bibinfo {year} {2012})},\ \Eprint
  {http://arxiv.org/abs/1112.2240} {arXiv:1112.2240 [astro-ph.HE]} \BibitemShut
  {NoStop}%
\bibitem [{\citenamefont {{He}}\ \emph {et~al.}(2012)\citenamefont {{He}},
  \citenamefont {{Liu}}, \citenamefont {{Wang}}, \citenamefont {{Nagataki}},
  \citenamefont {{Murase}},\ and\ \citenamefont {{Dai}}}]{He+12grbnu}%
  \BibitemOpen
  \bibfield  {author} {\bibinfo {author} {\bibfnamefont {H.-N.}\ \bibnamefont
  {{He}}}, \bibinfo {author} {\bibfnamefont {R.-Y.}\ \bibnamefont {{Liu}}},
  \bibinfo {author} {\bibfnamefont {X.-Y.}\ \bibnamefont {{Wang}}}, \bibinfo
  {author} {\bibfnamefont {S.}~\bibnamefont {{Nagataki}}}, \bibinfo {author}
  {\bibfnamefont {K.}~\bibnamefont {{Murase}}}, \ and\ \bibinfo {author}
  {\bibfnamefont {Z.-G.}\ \bibnamefont {{Dai}}},\ }\href {\doibase
  10.1088/0004-637X/752/1/29} {\bibfield  {journal} {\bibinfo  {journal}
  {\apj}\ }\textbf {\bibinfo {volume} {752}},\ \bibinfo {eid} {29} (\bibinfo
  {year} {2012})},\ \Eprint {http://arxiv.org/abs/1204.0857} {arXiv:1204.0857
  [astro-ph.HE]} \BibitemShut {NoStop}%
\end{thebibliography}
%

\end{document}